%% file: main.tex
\documentclass[11pt]{article}
\input{header}

\usepackage{microtype}
\newcommand{\wtO}{\widetilde O}
\newcommand{\E}{\mathbb E}
\newcommand{\Var}{\operatorname{Var}}
\newcommand{\norm}[1]{\left\lVert #1\right\rVert}
\newcommand{\ip}[2]{\langle #1,#2\rangle}

\newcommand{\bb}{\mathbf{b}}

\newcommand{\ve}{\mathbf e}
\newcommand{\vf}{\mathbf f}
\newcommand{\FOR}{\textsc{Fix Orientation Rule}}

\newcommand{\swagato}[1]{}
\newcommand{\sagnik}[1]{}

\title{A Simple Subquadratic Cut Query Protocol for Directed Reachability}
\author{Henry Fleischmann\thanks{Carnegie Mellon University, USA} \and Sagnik Mukhopadhyay\thanks{University of Birmingham, UK} \and Swagato Sanyal\thanks{University of Sheffield, UK}}
\date{}

\begin{document}
\maketitle

\thispagestyle{empty}
\pagenumbering{roman}

\input{abstract}

\clearpage
\bigskip
\setcounter{tocdepth}{2}
\tableofcontents

\clearpage

\pagenumbering{arabic}
\setcounter{page}{1}
\input{introduction}

\input{blueprint}
\section{Missing Proofs of \texorpdfstring{$\wtO(n^{9/5})$}{O-tilde(n to the power 9/5)} Query Protocol}\label{sec:warmup}
\label{sec:basic-part1}
In this section, we provide the technical details omitted
from the warm-up implementation of Step~2 in
Section~\ref{sec:blueprint}, giving a complete description
and analysis of this step. Recall that this step has two parts:
first, we construct a sequence of integral orientation
rules whose average imbalance is close to the target
$\bb^*$; then, we round this sequence to a single
orientation rule while controlling the additional error.

The first part is given by Algorithm~\ref{alg:simple-average}.
As discussed in Section~\ref{sec:approx-cut-implementation},
it is based on the follow-the-regularised-leader (FTRL)
framework in online optimisation. In the next subsection,
we introduce this framework and state the regret guarantee
that we will use to analyse the algorithm.

\subsection{An online optimisation black box}
Consider the following online optimisation game. In each round $t$, an algorithm first chooses a vector
$\by_t\in\mathbb R^n$. It then receives a loss vector
$\bl_t\in\mathbb R^n$ and incurs loss
$\ip{\bl_t}{\by_t}$.
The loss vector may depend on the algorithm's current choice
and the entire preceding history. The objective of the algorithm is to minimise the total loss incurred $\sum_{t=1}^T \ip{\bl_t}{\by_t}$.

For a fixed comparator $\bu\in\mathbb R^n$, define
the \emph{regret relative to $\bu$} by
\[
\Reg_T(\bu)
 :=
 \sum_{t=1}^T
 \ip{\bl_t}{\by_t}- \sum_{t=1}^T \ip{\bl_t}{\bu}.
\]
Intuitively, this captures how much the algorithm loses for not choosing $\bu$ as $\by_t$ in all the rounds.

Now consider the algorithm that starts with $\by_1=\mathbf 0$ and,
after receiving $\bl_t$, chooses
\[
 \by_{t+1}
 := \by_t-\eta\bl_t,
\]
where $\eta>0$ is a parameter.
This update rule is an instance of Euclidean FTRL;
see Section~3 of Trevisan's blog post~\cite{Trevisan19}
for an exposition. The following theorem bounds its
regret relative to any fixed comparator $\bu$.

\begin{theorem}[Euclidean FTRL, Theorem~2.4 of Shalev-Shwartz~\cite{SS12}]
\label{thm:ftrl}
For the above algorithm, every $\eta>0$, every sequence
of loss vectors, and every $\bu\in\mathbb R^n$,
\[
 \Reg_T(\bu)
 \leq
 \frac{\norm{\bu}_2^2}{2\eta}
 +\eta\sum_{t=1}^T\norm{\bl_t}_2^2.
\]
\end{theorem}

Theorem~\ref{thm:ftrl} is
Theorem~2.4 of Shalev-Shwartz~\cite{SS12}. Optimising its parameter gives
the following consequence, which we will invoke in our
application.

\begin{corollary}[Optimised regret bound]
\label{cor:ftrl-optimised}
Let $D,L>0$ and let $T\geq1$ be an integer.
Run the above algorithm with
\[
 \eta=\frac{D}{L\sqrt{2T}}.
\]
If $\norm{\bl_t}_2\leq L$ in every round,
then, simultaneously for every $\bu$ satisfying
$\norm{\bu}_2\leq D$,
\[
 \Reg_T(\bu)\leq DL\sqrt{2T}.
\]
\end{corollary}
\begin{proof}
By Theorem~\ref{thm:ftrl},
\[
 \Reg_T(\bu)
 \leq \frac{D^2}{2\eta}+\eta TL^2.
\]
For the stated choice of $\eta$, both terms equal
$DL\sqrt{T/2}$.
\end{proof}

The algorithm as stated here needs $D,L,T$ (to set $\eta$), but does not need the
comparator $\bu$ or the future loss vectors.
The guarantee holds for adaptively chosen loss vectors.
Since it holds simultaneously for all admissible
comparators, we may choose $\bu$ after observing
the entire sequence.

\subsection{Analysis of averaging and rounding}

We now complete the analysis of the two parts of the warm-up construction
from Section~\ref{sec:approx-cut-implementation}.
Throughout, $\bz_1,\ldots,\bz_T$ are the orientation rules
generated by Algorithm~\ref{alg:simple-average}, and
\[
 \bar{\bz}:=\frac1T\sum_{t=1}^T\bz_t.
\]
Since imbalance is linear in the fractional orientation rule,
\[
 \bb_{\bar{\bz}}
 =\frac1T\sum_{t=1}^T\bb_{\bz_t}.
\]

\paragraph{The averaging guarantee.}
We first prove Lemma~\ref{lem:simple-average} stated in Section~\ref{sec:approx-cut-implementation} that asserts that $\bb_{\bar{\bz}}$ is close to $\bb^*$.

\simpleaveraging*

Recall that Algorithm~\ref{alg:simple-average} uses
\[
 \bl_t=\bb^*-\bb_{\bz_t},
 \qquad
 \by_{t+1}=\by_t-\eta\bl_t,
 \qquad
 \eta=\frac{1}{2n\sqrt{2T}}.
\]
Thus the vectors $\by_t$ follow the FTRL update from
the preceding subsection. We first show that the algorithm incurs nonnegative loss
in every iteration (Claim~\ref{clm:warmup-nonnegative-loss}). We then choose a fixed comparator
whose cumulative loss is
$-T\norm{\bb_{\bar{\bz}}-\bb^*}_1$ (Claim~\ref{clm:warmup-comparator-loss}).
Comparing these two losses using the FTRL regret bound
will imply Lemma~\ref{lem:simple-average}

\begin{claim}[Nonnegative algorithm loss]
\label{clm:warmup-nonnegative-loss}
For every iteration $t$,
\[
 \ip{\bl_t}{\by_t}\geq0.
\]
\end{claim}

\begin{proof}
For any orientation $D$ of $H$,
\[
 \ip{\bb_D}{\by_t}
 =\sum_{(u,v)\in E(D)}
   \bigl(\by_t(u)-\by_t(v)\bigr).
\]
Indeed, an edge $u\to v$ contributes $+1$ to the
imbalance at $u$ and $-1$ to the imbalance at $v$.
Its contribution to this inner product is therefore
minimised by directing it from the endpoint with the
smaller $\by_t$-value to the endpoint with the larger
one. This is exactly how $\bz_t$ assigns directions.
Consequently,
\[
 \ip{\bb_{\bz_t}}{\by_t}
 =-\sum_{\{u,v\}\in E(H)}|\by_t(u)-\by_t(v)|
 \leq \ip{\bb_D}{\by_t}
 \qquad
 \text{for every orientation $D$ of $H$}.
\]
Since $\bb^*$ is feasible, some orientation of $H$
has imbalance $\bb^*$. Hence
\[
 \ip{\bl_t}{\by_t}
 =\ip{\bb^*}{\by_t}
  -\ip{\bb_{\bz_t}}{\by_t}
 \geq0.
\]
\end{proof}

The second claim identifies a comparator whose cumulative
loss measures the average imbalance error.

\begin{claim}[Comparator loss]
\label{clm:warmup-comparator-loss}
Define $\bu\in\{-1,+1\}^V$ by
\[
 \bu(v)=
 \begin{cases}
  +1,&\text{if }\bb^*(v)-\bb_{\bar{\bz}}(v)\leq0,\\
  -1,&\text{otherwise}.
 \end{cases}
\]
Then
\[
 \sum_{t=1}^T\ip{\bl_t}{\bu}
 =-T\norm{\bb_{\bar{\bz}}-\bb^*}_1.
\]
\end{claim}

\begin{proof}
By the definitions of $\bl_t$ and $\bar{\bz}$,
\begin{align*}
 \sum_{t=1}^T\ip{\bl_t}{\bu}
 &=\ip{\sum_{t=1}^T(\bb^*-\bb_{\bz_t})}{\bu}\\
 &=T\ip{\bb^*-\bb_{\bar{\bz}}}{\bu} \\
 &=-T\norm{\bb_{\bar{\bz}}-\bb^*}_1
\end{align*}
The last equality follows coordinate by coordinate:
$\bu(v)$ has the opposite sign to
$\bb^*(v)-\bb_{\bar{\bz}}(v)$ whenever this difference
is nonzero.
\end{proof}

We now combine the two claims to prove Lemma~\ref{lem:simple-average}.
\begin{proof}[Proof of Lemma~\ref{lem:simple-average}]
Choose $\bu$ as in Claim~\ref{clm:warmup-comparator-loss}.
By the definition of regret and the two claims,
\begin{align*}
 \Reg_T(\bu)
 &=\sum_{t=1}^T\ip{\bl_t}{\by_t}
   -\sum_{t=1}^T\ip{\bl_t}{\bu}\\
 &=\sum_{t=1}^T\ip{\bl_t}{\by_t}
   +T\norm{\bb_{\bar{\bz}}-\bb^*}_1&& \text{by Claim~\ref{clm:warmup-comparator-loss}}\\
 &\geq T\norm{\bb_{\bar{\bz}}-\bb^*}_1. && \text{by Claim~\ref{clm:warmup-nonnegative-loss}}
\end{align*}
We upper bound the regret using
Corollary~\ref{cor:ftrl-optimised}.
Every coordinate of $\bu$ has absolute value $1$.
Also, both $\bb^*(v)$ and $\bb_{\bz_t}(v)$ lie between
$-\deg_H(v)$ and $\deg_H(v)$, so
$|\bl_t(v)|\leq2n$. Therefore,
\[
 \norm{\bu}_2=\sqrt n,
 \qquad
 \norm{\bl_t}_2\leq \left(n\cdot(2n)^2\right)^{1/2} = 2n^{3/2}.
\]
For these bounds, the step size prescribed by Corollary~\ref{cor:ftrl-optimised} is
\[
 \frac{\sqrt n}{2n^{3/2}\sqrt{2T}}
 =\frac{1}{2n\sqrt{2T}},
\]
which is precisely the step size used by the algorithm.
Thus by Corollary~\ref{cor:ftrl-optimised} we have that
\[
 T\norm{\bb_{\bar{\bz}}-\bb^*}_1
 \leq\Reg_T(\bu)
 \leq \sqrt n \cdot (2n^{3/2})\cdot \sqrt{2T} = 2\sqrt2\,n^2\sqrt T.
\]
Dividing by $T$ gives
\[
 \norm{\bb_{\bar{\bz}}-\bb^*}_1
 \leq\frac{2\sqrt2\,n^2}{\sqrt T}
 \leq\frac{4n^2}{\sqrt T}.
\]

Finally, each iteration computes one imbalance vector
using $O(n)$ queries by Observation~\ref{obs:imbalance}.
All other operations require no queries, so the total
query complexity is $O(nT)$.
\end{proof}

\paragraph{The rounding guarantee.}
Recall that, independently for each unordered pair,
we choose an iteration uniformly from $\{1,\ldots,T\}$
and assign the direction prescribed in that iteration.
Let $\mathcal D$ be the resulting orientation rule
and $D$ its induced orientation of $H$.
We now bound the error introduced by this rounding.

\independentrounding*

\begin{proof}
For each vertex $v$, let $N_H(v)$ denote its set of
neighbours in $H$. For $w\in N_H(v)$, define
\[
 a_t(v,w)=
 \begin{cases}
  +1,&\text{if $\bz_t$ directs the pair from $v$ to $w$},\\
  -1,&\text{if $\bz_t$ directs the pair from $w$ to $v$}.
 \end{cases}
\]
Thus
\[
 \bb_{\bz_t}(v)=\sum_{w\in N_H(v)}a_t(v,w).
\]

The rounding procedure chooses an iteration
$I_{\{v,w\}}$ uniformly from $\{1,\ldots,T\}$,
independently for each unordered pair $\{v,w\}$.
Let
\[
 X_{v,w}:=a_{I_{\{v,w\}}}(v,w)
\]
be the contribution of the edge $\{v,w\}$ to the
imbalance at $v$ after rounding. Then
\[
 \bb_D(v)=\sum_{w\in N_H(v)}X_{v,w},
 \qquad
 \E[X_{v,w}]=\frac1T\sum_{t=1}^T a_t(v,w).
\]
Consequently,
\begin{align*}
 \E[\bb_D(v)]
 &=\sum_{w\in N_H(v)}\E[X_{v,w}]\\
 &=\frac1T\sum_{t=1}^T
       \sum_{w\in N_H(v)}a_t(v,w)\\
 &=\frac1T\sum_{t=1}^T\bb_{\bz_t}(v)
 =\bb_{\bar{\bz}}(v).
\end{align*}

For a fixed vertex $v$, the variables
$(X_{v,w})_{w\in N_H(v)}$ are independent, since they
depend on the choices for distinct unordered pairs.
Moreover, $X_{v,w}\in\{-1,+1\}$, so
\[
 \Var(X_{v,w})
 =\E[X_{v,w}^2]-\E[X_{v,w}]^2
 =1-\E[X_{v,w}]^2
 \leq1.
\]
It follows that
\[
 \Var(\bb_D(v))
 =\sum_{w\in N_H(v)}\Var(X_{v,w})
 \leq |N_H(v)|
 =\deg_H(v).
\]

We next bound the expected absolute error at $v$.
By Cauchy--Schwarz and the expectation identity above,
\begin{align*}
 \E\bigl[|\bb_D(v)-\bb_{\bar{\bz}}(v)|\bigr]
 &\leq
 \sqrt{\E\bigl[(\bb_D(v)-\bb_{\bar{\bz}}(v))^2\bigr]}\\
 &=\sqrt{\Var(\bb_D(v))}\\
 &\leq\sqrt{\deg_H(v)} \\
 &\leq \sqrt n.
\end{align*}
Summing these bounds over all vertices gives
\begin{align*}
 \E\bigl[\norm{\bb_D-\bb_{\bar{\bz}}}_1\bigr]
 &=\sum_{v\in V(H)}
   \E\bigl[|\bb_D(v)-\bb_{\bar{\bz}}(v)|\bigr]\\
 &\leq\sum_{v\in V(H)}\sqrt n\\
 &= n^{3/2}
 \leq\frac{n^2}{\sqrt T},
\end{align*}
where the last inequality uses $T\leq n$.
This step uses only linearity of expectation;
the imbalances at different vertices need not be
independent.

Finally, Markov's inequality yields
\[
 \Pr\left[
   \norm{\bb_D-\bb_{\bar{\bz}}}_1
   >\frac{4n^2}{\sqrt T}
 \right]
 \leq
 \frac{\E[\norm{\bb_D-\bb_{\bar{\bz}}}_1]}
      {4n^2/\sqrt T}
 \leq\frac14.
\]
Thus the claimed error bound holds with probability
at least $3/4$, and in particular at least $1/2$.
\end{proof}

\section{AI Disclosure} \label{sec:ai}
The algorithms were first discovered by ChatGPT 5.6 Sol and 6 Astra Pro using interaction. Before prompting the LLM, we were aware of the result \cite{CutReachability2026} and that it was generated using an AI harness. Our goal was to get a simpler and more `combinatorial' algorithm that achieves subquadratic query complexity. As such, we have fully rewritten the paper for human-centred exposition with a particular focus on understandability. We used additional AIs for editorial aspects. We take full responsibility for the manuscript’s content and the correctness of results. We do not claim any credit for achieving subquadratic complexity first.


\bibliographystyle{alpha}
{\small
\bibliography{references, biblio}
}

\clearpage
\appendix

\section{A Faster Implementation: \texorpdfstring{$\wtO(n^{5/3})$}{O-tilde(n to the power 5/3)} Queries}\label{sec:improved}

Throughout the appendix, the input is a simple undirected graph $H$
and a feasible target imbalance
vector $\bb^*$.  We improve the approximation step in two places. Optimistic mirror
descent reduces the average imbalance error from $O(n^2/\sqrt T)$ to
$O(n^2/T)$, and dependent rounding gives constant error at each vertex, thereby incurring a negligible additional error of at most $O(n)$. We will need one additional primitive not outlined in \Cref{sec:blueprint}.
\begin{lemma}[Weighted neighbour sums]\label{lem:weighted-count}
Given $v\in V$, $X\subseteq V\setminus\{v\}$, and known weights
$a_w\in[-1,1]$, we can approximate $\sum_{w\in X}c_{vw}a_w$
to additive error $n^{-c}$ using $O(\log n)$ undirected cut queries,
for any fixed $c>0$. Doing this at every vertex gives
$\ell_\infty$ error $n^{-c}$ using $O(n\log n)$ queries.
\end{lemma}
\begin{proof}
Truncate each weight to $O(\log n)$ binary digits with error at most
$n^{-c-1}$, changing the sum by at most $n^{-c}$.
For each digit, separately count the neighbours of $v$ in $X$
whose positive or negative weights have that digit nonzero in their magnitude.
Each count takes $O(1)$ queries by \Cref{obs:edge-counting}.
Subtract the negative count from the positive count, multiply by
the digit's value, and sum over the $O(\log n)$ digits.
This computes the truncated weighted sum with the claimed query bound.
\end{proof}

\subsection{An online optimisation black box with predictions}
\label{sec:optimistic-regret}

We use a second standard regret bound, this time for an algorithm
that receives a prediction of the loss vector before making its
choice. Consider the online game from the warm-up, with choices
restricted to $\calB=[-1,1]^{d}$, with $d \leq 2n^2$.
Let $P$ restrict each coordinate to $[-1,1]$.
Fix $\eta=1/8$, a horizon $1\leq T\leq n$, and initialise
$\bv_1=\mathbf 0$. In each round $t$, do the following:
\begin{enumerate}
\item Receive a prediction $\widehat{\bl}_t$ of the next loss vector,
      and choose the trial vector
      \[
       \bv_t^+=P(\bv_t-\eta\widehat{\bl}_t).
      \]
\item Receive the actual loss vector $\bl_t$ and incur loss
      $\ip{\bl_t}{\bv_t^+}$.
\item Set the vector for the next round to
      \[
       \bv_{t+1}=P(\bv_t-\eta\bl_t).
      \]
\end{enumerate}
Both updates start from the previous round vector $\bv_t$.
This algorithm is called \emph{optimistic mirror descent}; its
prediction is the reason for the adjective ``optimistic''.
For a comparator $\bu\in\calB$, define regret as before:
\[
 \Reg_T(\bu)=\sum_{t=1}^T\ip{\bl_t}{\bv_t^+-\bu}.
\]
We use the following bound, specialised to our parameters.

\begin{theorem}[{Rakhlin and Sridharan~\cite[Lemma~1, equation~(2)]{RS13}}]
\label{thm:optimistic-regret}
Suppose that in every round
\begin{equation}\label{eq:prediction-error}
 \norm{\bl_t-\widehat{\bl}_t}_2
 \leq2\norm{\bv_t^+-\bv_t}_2+\frac2{n^2}.
\end{equation}
Then, simultaneously for every $\bu\in\calB$,
\[
 \Reg_T(\bu)\leq O(n^2).
\]
\end{theorem}

Condition~\eqref{eq:prediction-error} requires the predicted loss
$\widehat{\bl}_t$ to be close to the actual loss $\bl_t$ when the
trial vector is close to the previous round vector. The predictions
and losses may depend on previous rounds, and the loss may depend
on the current choice. The guarantee holds for all comparators at
once, so we may choose the comparator afterwards.

\subsection{Averaging fractional orientations}\label{sec:fast-average}

As in the warm-up, we choose the losses so that a regret bound
implies a bound on imbalance. The state consists of a fractional
orientation rule $\bz\in[-1,1]^{\binom n2}$ and vertex
potentials $\bx\in[-1,1]^n$. Positive excess outflow raises a
vertex's potential, encouraging subsequent edge directions to point
towards it. Unlike in the warm-up, the fractional orientation is now itself part of the
online vector.

\paragraph{Losses and updates.}
To express the algorithm in the online framework, let
$\bv(\bz,\bx)$ have one coordinate per edge of $H$, carrying
its pair's value in $\bz$, and $n$ coordinates per vertex $v$,
each carrying $\bx(v)$. Thus, $\bv \in \calB$, for $d = |E(H)| + n^2 \leq 2 n^2$. Unlike in the warm-up, each round may now involve a truly fractional orientation $\bz$, and we cannot efficiently compute the induced imbalance vector $\bb_{\bz}$ exactly. Instead, we use an approximation $\widetilde{\bb}_{\bz}$ matching on each entry up to $O(\log n)$ bits of precision, computed by applying \Cref{lem:weighted-count}.
With this in mind, define the loss vector at state $(\bz,\bx)$ by
\[
 \begin{aligned}
 \bl(\bz,\bx)(e)&=\bx(u)-\bx(v)
 &&(e=\{u,v\},\ u<v),\\
 \bl(\bz,\bx)(v,i)&=\frac{\bb^*(v)-\widetilde{\bb}_{\bz}(v)}n
 &&(1\leq i\leq n).
 \end{aligned}
\]
The first formula can be evaluated on every potential pair. The $n$ coordinates for a
vertex evolve identically. Hence the implementation keeps only
$\bz$ and $\bx$; it need not learn the edges or store repeated coordinates.
Each loss evaluation to $b = O(\log n)$ bits of precision uses $O(n\log n)$ queries by
\Cref{lem:weighted-count}.

Fix $\eta=1/8$, an integer $1\leq T\leq n$, and initialise
$\bz_1=\mathbf 0$ and $\bx_1=\mathbf 0$.
In round $t$, write $\bv_t=\bv(\bz_t,\bx_t)$ and set
\begin{equation}\label{eq:trial}
 \widehat{\bl}_t=\bl(\bz_t,\bx_t),\qquad
 \bv_t^+=P(\bv_t-\eta\widehat{\bl}_t)
       =\bv(\bz_t^+,\bx_t^+).
\end{equation}
Then evaluate the loss at the trial state and update from $\bv_t$:
\begin{equation}\label{eq:correction}
 \bl_t=\bl(\bz_t^+,\bx_t^+),\qquad
 \bv_{t+1}=P(\bv_t-\eta\bl_t)
          =\bv(\bz_{t+1},\bx_{t+1}).
\end{equation}
Apply the edge updates to all potential pairs. After $T$ rounds,
return $\bar{\bz}=T^{-1}\sum_{t=1}^T\bz_t^+$.
These are the extragradient updates of Mirror-Prox~\cite{Nemirovski};
we analyse them using Theorem~\ref{thm:optimistic-regret}.

\begin{lemma}[Faster averaging]\label{lem:fast-average}
For every integer $1\leq T\leq n$, the above algorithm uses
$O(nT\log n)$ queries and returns a fractional orientation rule
$\bar{\bz}$ with $O(\log n)$-bit rational probabilities and
\[
 \norm{\bb_{\bar{\bz}}-\bb^*}_1=O(n^2/T).
\]
\end{lemma}
\begin{proof}
There are two loss evaluations per round, giving the query bound.
The updates keep all coordinates
on the grid $1/(64n2^b)$. Averaging and converting to probabilities
adds a denominator factor of $O(T)$, so $T\leq n$ gives the claimed bit bound.

We first explain how the regret bound implies the imbalance bound.
Choose $\bh\in\{-1,1\}^V$ to have the signs of
$\bb_{\bar{\bz}}-\bb^*$, taking $+1$ at zero. By feasibility, there is an orientation
$D^*$ of $H$ with imbalance $\bb^*$. Let $\bu$ use its direction
on each edge and repeat $\bh(v)$ on the $n$ coordinates for
vertex $v$. The edge and vertex contributions to regret give
\begin{align*}
 \ip{\bl_t}{\bv_t^+-\bu}
 &=\ip{\bx_t^+}{\bb_{\bz_t^+}-\bb^*}
   +\ip{\bx_t^+-\bh}{\bb^*-\widetilde{\bb}_{\bz_t^+}}\\
 &=\ip{\bh}{\bb_{\bz_t^+}-\bb^*}
   +\ip{\bx_t^+-\bh}{\bb_{\bz_t^+}-\widetilde{\bb}_{\bz_t^+}}.
\end{align*}
The final term has absolute value at most $1/n$ given a sufficiently strong approximation of the imbalance (and this is easy to achieve by just increasing the bit precision of the estimate). Since $T\leq n$, summing and using
the choice of $\bh$ gives
\begin{equation}\label{eq:fast-regret-error}
 T\norm{\bb_{\bar{\bz}}-\bb^*}_1
 \leq\Reg_T(\bu)+1\leq O(n^2),
\end{equation}
once Theorem~\ref{thm:optimistic-regret} applies.

It remains to check the Lipschitz estimate
\eqref{eq:prediction-error}. Write
$\Delta\bx=\bx_t^+-\bx_t$ and $\Delta\bz=\bz_t^+-\bz_t$.
The degree bound $\deg_H(v)\leq n$ and Cauchy--Schwarz give
\begin{align*}
 \sum_{e=\{u,v\}\in E(H)}
   (\Delta\bx(u)-\Delta\bx(v))^2
 &\leq 2\sum_{v\in V}\deg_H(v)\Delta\bx(v)^2\\
 &\leq 2n\norm{\Delta\bx}_2^2,
 \\[1ex]
\norm{\bb_{\bz_t^+}-\bb_{\bz_t}}_2^2
 &= \sum_{v\in V}
    \left(
      \sum_{\substack{w>v\\\{v,w\}\in E(H)}}\Delta\bz(v,w)
      -
      \sum_{\substack{w<v\\\{w,v\}\in E(H)}}\Delta\bz(w,v)
    \right)^2\\
 &\leq \sum_{v\in V}\deg_H(v)
          \sum_{\substack{e\in E(H)\\e\ni v}}\Delta\bz(e)^2\\
 &\leq n\sum_{v\in V}
          \sum_{\substack{e\in E(H)\\e\ni v}}\Delta\bz(e)^2\\
 &=2n\sum_{e\in E(H)}\Delta\bz(e)^2.
\end{align*}
Here the first bound uses $(a-b)^2\leq 2a^2+2b^2$.
For the second, apply Cauchy--Schwarz to the signed sum of
incident edge changes at each vertex; the final equality
counts each edge at its two endpoints. Each vertex loss is divided by
$n$ and repeated $n$ times, so with exact imbalances these estimates
bound the squared loss difference by
$2(n\norm{\Delta\bx}_2^2+\sum_e\Delta\bz(e)^2)
=2\norm{\bv_t^+-\bv_t}_2^2$. However, technically $\bl_t$ and $\widehat{\bl}_t$ are computed only via strong estimates of the imbalances $\bb_{\bz_t}$ and $\bb_{\bz_t^+}$. Nonetheless, by choosing sufficiently large bit precision $b$, we may assume that
\[
\|\widetilde{\bb}_{\bz_t^+} - \widetilde{\bb}_{\bz_t}\|_2 \leq \|\widetilde{\bb}_{\bz_t^+} - \bb_{\bz_t^+}\|_2 + \|\bb_{\bz_t^+} - \bb_{\bz_t} \|_2 + \|\bb_{\bz_t} - \widetilde{\bb}_{\bz_t}\|_2 \leq \|\bb_{\bz_t^+} - \bb_{\bz_t} \|_2 + 1/\poly(n),
\]
using the triangle inequality. Hence, using this additive error bound and the fact that $\|\bb_{\bz_t^+} - \bb_{\bz_t} \|_2 \leq n^{3/2}$, we get an analogous bound as above on $\|\widetilde{\bb}_{\bz_t^+} - \widetilde{\bb}_{\bz_t}\|_2^2$, up to an additive error of at most $1/\poly(n)$.
Hence, \eqref{eq:prediction-error} holds and \Cref{thm:optimistic-regret}  applies, as required.
\end{proof}

\subsection{Rounding one bit at a time}
The second improvement over the warm-up algorithm is a more efficient rounding procedure. Let us build up to this rounding procedure with some intuition.

\begin{figure}[htbp]
\centering
\includegraphics[width=\linewidth]{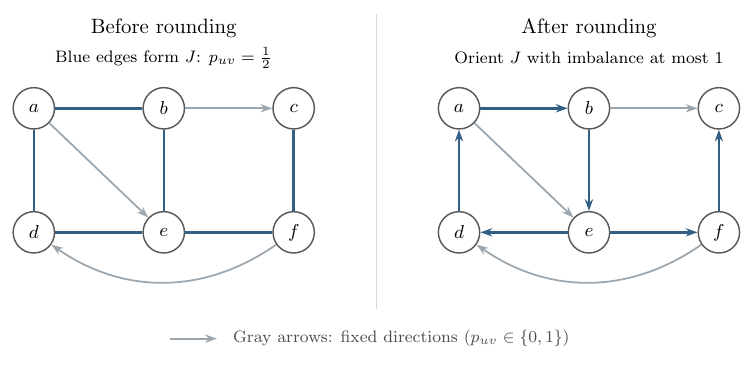}
\caption{The toy case of rounding probabilities in $\{0,1/2,1\}$.}
\label{fig:toy-rounding}
\end{figure}

First, note that each edge weight induced by the fractional orientation rule $\bar{\bz}$ can be interpreted as a probability via the map $p_{uv}=(1+\bar{\bz}(u,v))/2$. This interpretation is well-motivated---the imbalance $\bb_{\bar{\bz}}$ is also the expected imbalance upon orienting each edge $(u,v)$ towards $v$ with probability $p_{uv}$. Suppose that these probabilities were all $0, 1/2, $ or $1$. How would we round to an appropriate integral orientation matching the vertex imbalances in this toy case? Well, the only edges requiring rounding are those with corresponding probabilities $1/2$. Call the graph on these edges $J$. Moreover, notice that these edges in $J$ contribute $0$ in expectation to the imbalances of their incident vertices. Crucially, this means that our goal is to compute an integral orientation of $J$ with imbalance $0$. See \Cref{fig:toy-rounding}.
In reality, computing an orientation with imbalance $0$ might not be possible, but luckily we can tolerate some error in our rounding procedure.

What about the general case with more varied probabilities? In fact, as we will show, the same procedure works, proceeding one bit at a time in the binary representations of the edge probabilities. Instead of orienting the edges in each rounding step, we round the least significant remaining bit in the probability. To implement this idea, we need a subroutine finding an integral orientation with imbalance close to $0$. \Cref{lem:balanced-2} gives exactly the desired guarantees.

\begin{lemma}[Balanced orientation]\label{lem:balanced-2}
Let $J$ be an undirected graph accessible by cut queries. There exists an always-correct randomised algorithm computing an orientation rule $\calD$ inducing an implicit orientation $D$ for $J$ with $\|\bb_D\|_{\infty} \leq 1.$
The algorithm uses $\wtO(n^{5/3})$ queries in expectation and with
high probability. The algorithm also only uses neighbor searches, edge tests, and vertex imbalance evaluations.
\end{lemma}
We defer the proof of \Cref{lem:balanced-2} to \Cref{app:balanced-completion}. Note that we can efficiently simulate neighbor searches, edge tests, and vertex imbalance evaluations using queries to $H$. With this subroutine in hand, we can implement the dependent rounding algorithm.

\begin{lemma}[Dependent rounding]\label{lem:dependent-round}
Let $\bz$ be a fractional orientation rule.
An always-correct randomised algorithm rounds $\bz$ to an orientation rule $\calD$ such that the induced orientation $D$ satisfies
\[
 \norm{\bb_D-\bb_{\bz}}_\infty=O(1).
\]
It uses $\wtO(n^{5/3})$ queries in expectation and with high
probability.
\end{lemma}
\begin{proof}
As in the proof sketch, we interpret the edge weights from $\bz$ as probabilities: $p_{uv}=(1+\bz(u,v))/2$. First round these probabilities down
to multiples of $2^{-B}$, where $B=\lceil\log_2 n\rceil$.
This changes each vertex's imbalance by $O(n2^{-B})=O(1)$. After this rounding, the probabilities correspond to binary strings on at most $B$ bits.

Starting from the least significant bit of the probabilities (at most the $B$th bit), consider the pairs whose current $j$th bit is nonzero. Denote the resultant subgraph (of the complete graph) by $F_j$. Apply \Cref{lem:balanced-2} to $F_j$, emulating the neighborhood searches, edge tests, and vertex imbalance evaluations by restricting only to the set of possible edges.
Let $\sigma_{uv}\in\{-1,1\}$ be the resulting orientation. On each edge in $F_j$, set
\[
 p_{uv}\leftarrow p_{uv}+2^{-j}\sigma_{uv}.
\]
That is, if the edge is oriented from $u$ to $v$ in the orientation outputted by \Cref{lem:balanced-2}, the $j$th bit is rounded up. Otherwise it is rounded down.

Let $D_j$ be the orientation used at this step. The change
in fractional imbalance is $2^{1-j}\bb_{D_j}$.
By \Cref{lem:balanced-2}, $D_j$ is almost perfectly balanced; for all $v$, $|\bb_{D_j}(v)|\leq 1$. Hence, since
$\sum_{j\geq1}2^{1-j}=2$, the total additional error is
$O(1)$ at each vertex. The final probabilities are integral.

Finally, there are $O(\log n)$ calls to Lemma~\ref{lem:balanced-2}, each
costing $\wtO(n^{5/3})$ queries. The probability guarantees are inherited from \Cref{lem:balanced-2}.
\end{proof}

\subsection{Putting the faster implementation together}
We can now complete the proof of \Cref{thm:main}.
\begin{proof}[Proof of Theorem~\ref{thm:main}]
By Lemma~\ref{lem:directed-to-undirected}, it suffices to construct
an orientation rule of $H$ matching
$\bb^*$. Apply the faster averaging algorithm \Cref{lem:fast-average},
followed by dependent rounding \Cref{lem:dependent-round}. This gives an orientation rule $\calD$ with induced orientation $D$ satisfying
\begin{equation}\label{eq:fast-accepted}
 \norm{\bb_D-\bb^*}_1
 \leq\norm{\bb_{\bar{\bz}}-\bb^*}_1+\norm{\bb_D-\bb_{\bar{\bz}}}_1
 =O(n^2/T+n)=O(n^2/T).
\end{equation}
Computing $\calD$ requires $\wtO(nT+n^{5/3})$ total queries including the
$O(n)$ queries needed to compute $\bb_D$ exactly. To complete the argument, we use the same flow subroutine as before; \Cref{lem:repair} guarantees the required imbalance and takes $\wtO(n^{5/3} + n^2/\sqrt{T})$ queries.
The total number of undirected cut queries is
\begin{equation}\label{eq:improved-tradeoff}
 \wtO\left(nT+\frac{n^2}{\sqrt T}+n^{5/3}\right)
 =\wtO(n^{5/3}).
\end{equation}
The result then follows by setting $T = n^{2/3}$; the probabilistic guarantees are inherited from the analogous guarantees of the subroutines.
\end{proof}

\section{Balanced Orientations by Completing a Partial Orientation}
\label{app:balanced-completion}

We now prove \Cref{lem:balanced-2}. The strategy is to reserve a small set of known edges such that if we round the remaining edges uniformly at random, we are very likely able to orient the known edges to balance the full orientation. The reserved edges are a union of $\wtO(n^{2/3})$ greedily computed spanning forests along with up to $\wtO(n^{5/3})$ additional randomly sampled edges.

\subsection{Completing a partial orientation}
We first show simple sufficient conditions for the reserved set of edges to be able to correct the imbalance incurred by orienting the other edges. The following basic observation is quite useful.
\begin{observation}[Existence of orientations]\label{obs:orientation}
Every undirected graph admits an orientation with imbalance at
most one at every vertex.
\end{observation}
\begin{proof}
Add an auxiliary vertex adjacent to every odd-degree vertex.
All degrees are now even, including the auxiliary vertex's degree,
since the number of odd-degree vertices is even. Compute an Eulerian tour in each component to direct all edges, ensuring equal in and out degrees. Finally, remove the auxiliary node and its incident edges. This changes the imbalance of each original vertex by at most $1$.
\end{proof}

Given \Cref{obs:orientation}, we can give a max flow based sufficient condition for the reserved edges.
\begin{lemma}[Completing a partial orientation]
\label{lem:balanced-completion}
Let $R$ be the graph of reserved edges of $J$, and let $D_0$
orient all other edges. If
\begin{equation}\label{eq:balanced-completion-cuts}
 \left|\sum_{v\in S}\bb_{D_0}(v)\right|\leq C_R(S)
 \qquad\text{for every }S\subseteq V,
\end{equation}
then $D_0$ can be extended to an orientation $D$ with imbalance at
most one at every vertex. Computing $D$ from $D_0$ uses only the explicit
graph $R$ and the vector $\bb_{D_0}$.
\end{lemma}
\begin{proof}
Consider the max flow problem on $R$ with vertex demands $-\bb_{D_0}$. By the max-flow mincut theorem, \Cref{eq:balanced-completion-cuts} implies that $-\bb_{D_0}$ is feasible. Indeed, since capacities and demands are integral, there is a feasible \emph{integral} flow.

To extend $D_0$ to $D$, first orient every edge carrying flow along the flow paths. Then, $D_0$, combined with the orientations of the edges from the flow, yields imbalance $0$ for each vertex. The last detail is to orient the edges in $R$ unused in the flow; we can orient this subgraph using \Cref{obs:orientation}, resulting in a final orientation $D$ with each vertex imbalance at most $1$.
\end{proof}
Given \Cref{lem:balanced-completion}, we know that it suffices to select a set of reserved edges $R$ of $J$ to learn explicitly such that \Cref{eq:balanced-completion-cuts} ultimately holds. Then, after computing $D_0$, we can extend it to $D$ as required for \Cref{lem:balanced-2}.

\subsection{Constructing a suitable reserve}
We now demonstrate how to construct an appropriate reserve edge set $R$. Set
$k=n^{2/3}$ and greedily compute a union of $10k\log n$ successive edge-disjoint spanning forests of $J$. Call the resultant subgraph $F$. Each spanning forest takes
$O(n\log n)$ cut queries by Lemma~\ref{lem:subgraph-operations},
excluding the pairs already removed from later neighbour searches.
Thus learning $F$ costs $\wtO(nk)$ queries. The subgraph $F$ is the Nagamochi-Ibaraki certificate \cite{nagamochi1992linear} and is sometimes called a $k$-connectivity certificate \cite{bodwin23}. $F$ naturally preserves various connectivity properties of $J$. Of relevance to our application, Bencz\'{u}r and Karger show the following property.
\begin{theorem}[{\cite[Section~8.2 and
Theorem~8.3]{BK}}] \label{thm:weak-edges}
$F$ induces a partition $\calP$ of $J$ such that each component is either $k$-edge connected or a singleton, and all inter-component edges belong to $F$.
\end{theorem}

In addition to including $F$ in $R$, we also sample each possible edge with probability $\Theta(\log n/ \sqrt{k})$. For each sampled candidate edge, we check whether it exists in $J$ in $O(1)$ cut queries and, if it does, add it to $R$. By a Chernoff bound, with high probability this takes $\widetilde{O}(n^2/\sqrt{k})$ total cut queries. This completes the construction of $R$. For all remaining edges we orient them independently and uniformly at random, giving the partial orientation $D_0$. The total query bound in constructing $R$ is $\widetilde{O}(n^2/\sqrt{k} + nk) = \widetilde{O}(n^{5/3})$ as needed.

To complete the proof of \Cref{lem:balanced-2}, it remains to show that $D_0$ and $R$ satisfy the requirements of \Cref{lem:balanced-completion} with high probability. (To get the always correct guarantee, we simply compute $\bb_{D}$ in $\wtO(n)$ queries and repeat if it does not satisfy $\|\bb_{D}\|_{\infty} \leq 1$.)

\begin{claim} \label{cla:completable-cond}
The reserved set $R$ and partial orientation $D_0$ satisfy the condition of \Cref{lem:balanced-completion} with high probability.
\end{claim}
To prove \Cref{cla:completable-cond}, the following classic cut-counting bound will be useful.

\begin{theorem}[Cut counting~{\cite[Theorem~8.4]{KS96}}]
\label{thm:balanced-cut-counting}
In a graph on at most $n$ vertices with minimum cut at least $\lambda$,
the number of cuts of size less than $\alpha \lambda$ is at most
$n^{O(\alpha)}$, for every integer $\alpha \geq 2$.
\end{theorem}
\begin{proof}[Proof of \Cref{cla:completable-cond}]
Recall that we need to lower bound the probability that \begin{equation*}
 \left|\sum_{v\in S}\bb_{D_0}(v)\right|\leq C_R(S)
 \qquad\text{for every }S\subseteq V.
\end{equation*}
We exploit the partition induced by $F$ from \Cref{thm:weak-edges}. Fix a non-singleton component $C \in \calP$ from the partition. Consider a nontrivial cut $\emptyset \subsetneq S_C \subsetneq C$, and suppose there are $a$ edges crossing this cut within $C$. That is, $|E_{J}(S_C, C \setminus S_C)| = a$. Let $u_{C}^-$ and $u_{C}^+$ be the numbers of unreserved edges crossing into and out of $S_C$ within $C$, respectively.  We show that, with very high probability, $\max\{u_{C}^-, u_{C}^+\} \leq a/2$.

To this end, recall that, in addition to $F$, $R$ is formed by sampling each edge with probability $\Theta(\log n / \sqrt{k})$. Then, by a Chernoff bound, since the expected number of edges in either direction is at most $a/2 - \Theta(a\log n/\sqrt{k})$, the probability that either direction has at least $a/2$ edges is at most $n^{-\Omega(a \log n /k)}$.
Group the internal cuts by
$(\alpha-1)k\leq a<\alpha k$, for integers
$\alpha\geq2$. By \Cref{thm:balanced-cut-counting}, since the component is $k$-edge connected, each group
contains at most $n^{O(\alpha)}$ cuts. Then, a union bound over these groups
and the at most $n$ parts gives total failure probability at most
\[
n\sum_{\alpha\geq2}n^{O(\alpha)} n^{-\Omega(\alpha \log n)}
 \leq n^{-c},
\]
for any fixed $c>0$, setting constants appropriately.

So, with high probability, for all non-trivial $S \subseteq V$, for all $C$ for which $S \cap C$ is non-trivial, the cut $S_C$ induced by $S$ in $C$ has at most half of its crossing edges in either direction.
Every edge between different parts is reserved, so
$\sum_{v\in S}\bb_{D_0}(v)$ is the sum of the signed imbalances
of the internal cuts induced by $S$ in the parts.
For each induced cut $S_C$, if there are $a$ crossing edges and $r$ of these are reserved, then $a = r + u_{C}^+ + u_{C}^-$, where $u_{C}^+, u_{C}^- \leq a/2$. In particular, this means that the magnitude of the imbalance contributed by $S_C$, $|u_{C}^+ - u_{C}^-|$, is at most
\[
|u_{C}^+ - u_{C}^-| = 2 \max\{u_{C}^+, u_{C}^-\} - (a - r) \leq a - (a-r) = r.
\]
Summing over all induced cuts $S_C$, this gives $\sum_{v\in S}\bb_{D_0}(v) \leq C_R(S)$. Thus the required condition
holds simultaneously for every $S$ with high probability.
\end{proof}

\section{The Flow Subroutine}\label{app:flow}

We are given undirected cut-query access to a simple graph $H$ on
$n$ vertices and a feasible target
imbalance vector $\bb^*$. Feasibility means that some orientation
$D^*$ of $H$ has imbalance $\bb^*$. We are also given an orientation rule $\mathcal D$ with the orientation $D$ that it induces on $H$ satisfying $\norm{\bb_D-\bb^*}_1=\Delta$. The objective is to find an orientation rule $\mathcal F$
whose induced orientation $F$ of $H$ satisfies
$\bb_F=\bb^*$. We call this problem \FOR.
Our goal is to prove the following lemma.
\begin{lemma}
\label{lem:flowmain}
There is a deterministic algorithm that solves \FOR\ using
$\wtO(n^{5/3}+n\sqrt{\Delta})$ undirected cut queries to $H$.
\end{lemma}

We prove Lemma~\ref{lem:flowmain} by formulating \FOR\ as a
network flow problem. Define the demand vector
\[
 \ve:=\frac{\bb_D-\bb^*}{2}.
\]
Its coordinates are integers because
$\ve(v)=\deg_{D}^+(v)-\deg_{D^*}^+(v)$. Every orientation has total
imbalance zero, since each edge contributes once with each sign.
Consequently, $\sum_v \ve(v)=0$, and the total supply is
\[
 K:=\sum_{v:\ve(v)>0}\ve(v)
   =\frac12\norm{\ve}_1
   =\frac{\Delta}{4}.
\]
The network $N$ has vertex set $V(H)\cup\{p,q\}$, where $p$ and
$q$ are fresh vertices, and the following arcs and capacities:
\begin{enumerate}
\item Each edge of $H$ appears as an arc of capacity one,
      directed according to $\mathcal D$.
\item For each $v\in V(H)$ with $\ve(v)>0$, there is an arc
      $p\to v$ of capacity $\ve(v)$.
\item For each $v\in V(H)$ with $\ve(v)<0$, there is an arc
      $v\to q$ of capacity $-\ve(v)$.
\end{enumerate}
The source and sink of $N$ are $p$ and $q$, respectively.
The following lemma gives its maximum flow value and query complexity.
\begin{lemma}
\label{lem:repair-maxflow}
Let $\Delta=\norm{\bb_D-\bb^*}_1$.
\begin{enumerate}
\item[(a)] The maximum $p$--$q$ flow value in $N$ is exactly
$\Delta/4$.
\item[(b)] An integral maximum flow can be computed explicitly
by a deterministic algorithm using
$\wtO(n^{5/3}+n\sqrt{\Delta})$
undirected cut queries to $H$.
\end{enumerate}
\end{lemma}
First, we finish the proof of
Lemma~\ref{lem:flowmain} assuming Lemma~\ref{lem:repair-maxflow}. Then we prove part (a) of Lemma~\ref{lem:repair-maxflow}. Its part (b) is an adaptation of the algorithm by \cite{ASW} to our setting and will be proved in the next subsection.
\begin{proof}[Proof of Lemma~\ref{lem:flowmain}]
Compute $\bb_D$ using $O(n)$ undirected cut queries, which also
determines $\Delta=\norm{\bb_D-\bb^*}_1$, every $\ve(v)$, and the
terminal arcs of $N$. The internal arcs remain implicit.
By Lemma~\ref{lem:repair-maxflow}, we can compute an explicit
integral $p$--$q$ flow $f$ in $N$ of value $\Delta/4$ using
$\wtO(n^{5/3}+n\sqrt{\Delta})$ undirected cut queries.

The total capacity leaving $p$ and the total capacity entering $q$
both equal $\Delta/4$. Thus every arc leaving $p$ or entering $q$
is saturated. By flow conservation, the net outgoing flow on the
internal arcs at each vertex $v$ is $\ve(v)$: the source arc supplies
$\ve(v)$ when it is positive, and the sink arc removes $-\ve(v)$ when it
is negative.

Every internal arc has capacity $1$, so its integral flow
is either $0$ or $1$. Define the orientation rule $\mathcal F$
as follows: reverse the direction prescribed by
$\mathcal D$ on every pair whose internal arc carries
one unit of flow, and retain its direction on every
other pair. Let $F$ be the orientation of $H$ induced
by $\mathcal F$.

Each reversed outgoing edge decreases the imbalance
by two, and each reversed incoming edge increases it
by two. Consequently,
\[
 \bb_F(v)=\bb_D(v)-2\ve(v)=\bb^*(v)
 \qquad\text{for every }v\in V(H).
\]
Since $\mathcal D$ and the flow are explicitly available,
the direction prescribed by $\mathcal F$ on any pair can be
evaluated without further queries. The total query complexity is therefore
$\wtO(n^{5/3}+n\sqrt{\Delta})$.
\end{proof}

Next, we prove part (a) of Lemma~\ref{lem:repair-maxflow}.
\begin{proof}[Proof of Lemma~\ref{lem:repair-maxflow} part (a)]
The total capacity leaving $p$ is
\[
 \sum_{v:\ve(v)>0}\ve(v)
 =\frac12\sum_v|\ve(v)|
 =\frac14\norm{\bb_D-\bb^*}_1
 =\frac{\Delta}{4},
\]
so the maximum flow value is at most $\Delta/4$.

For the reverse inequality, fix an orientation $D^*$ of $H$ with
imbalance $\bb^*$, which exists by feasibility. On each internal
arc of $N$, send one unit of flow if its direction in $\mathcal D$ differs
from its direction in $D^*$, and zero otherwise. At each vertex
$v$, the net outgoing flow on these arcs is
\[
 \frac{\bb_D(v)-\bb_{D^*}(v)}2=\ve(v).
\]
Saturate every terminal arc. If $\ve(v)>0$, the arc $p\to v$ supplies
the excess outgoing flow $\ve(v)$; if $\ve(v)<0$, the arc $v\to q$
removes the excess incoming flow $-\ve(v)$. Thus flow conservation
holds at every internal vertex, and all capacities are respected.
All arcs leaving $p$ are saturated, so this flow has value
$\sum_{v:\ve(v)>0}\ve(v)=\Delta/4$.

\end{proof}
\subsection{Proof of Lemma~\ref{lem:repair-maxflow} part (b)}
 Recall that we have undirected cut query access to $H$, a target imbalance vector $\bb^*$, and an orientation rule $\mathcal D$ with imbalance vector $\bb_D$ of the induced orientation $D$ of $H$ satisfying $\norm{\bb_D-\bb^*}_1=\Delta$. These define the $p$--$q$ maxflow instance $N$. We will implement Dinic's algorithm in our setting to compute a $p$--$q$ maximum flow in the network $N$.
 \subsubsection{Proof overview}

Our proof adapts the implementation of Dinic's algorithm
by Anand, Saranurak, and Wang~\cite{ASW} to our setting.
Dinic's algorithm proceeds in phases, each of which
increases the current flow by sending additional flow
along shortest source--sink paths in the residual network.
We recall the relevant definitions in the subsequent sections.

The main result we adapt is \cite[Claim~4.3]{ASW},
which implements one such phase efficiently.
Two observations allow us to use it here.
\begin{enumerate}
\item Although their original network is undirected,
the procedure operates on its directed residual network.
The proof also applies when the original network is
directed, as ours is.

\item To search the residual network, the procedure
only needs to test whether a given vertex $u$ has an
outgoing residual arc to a given set $B$.
These tests are called \emph{singleton BIS queries}.
We implement each test, and each numerical
residual-capacity query needed to send flow, using
$O(1)$ cut queries
(Lemma~\ref{lem:residual-oracles}).
\end{enumerate}

The remaining analysis is a specialisation of standard
arguments for Dinic's algorithm to our network $N$,
following the BFS-layer analysis in
\cite[Section~4.2]{ASW}. We account for the unit capacities
and simplicity of the internal graph, the larger
capacities on terminal arcs, and the required flow value
$K=\Delta/4$. These arguments yield the query bound
$\wtO(n^{5/3}+n\sqrt{\Delta})$.
 \subsubsection{Blocking flows and Dinic's algorithm}

Dinic's algorithm constructs a maximum flow by repeatedly
sending additional flow along currently shortest residual
paths. We recall the relevant definitions; see also
\cite[Section~4.2]{ASW}.

\begin{definition}[Residual network]
Let $f$ be a feasible flow in a network $N$ with source $p$
and sink $q$. For each arc $u\to v$ of capacity $c$ carrying
flow $x$, the residual network $N_f$ has a forward arc
$u\to v$ of capacity $c-x$ and a reverse arc $v\to u$ of
capacity $x$. Zero-capacity arcs are omitted.
\end{definition}
For every ordered pair of distinct vertices $(u,v)$,
let $c(u,v)$ and $f(u,v)$ denote its original capacity
and flow, taking both to be zero if $u\to v$ is absent
from $N$. Let $r_f(u,v)$ denote its residual capacity,
also taking this to be zero if the residual arc is absent.

Our network has at most one original arc between
any pair of vertices. If this arc is $u\to v$, its
forward residual capacity is $c(u,v)-f(u,v)$.
If instead the original arc is $v\to u$, its reverse
residual arc points from $u$ to $v$ and has capacity
$f(v,u)$. If neither original arc exists, the residual
capacity is zero. These cases are expressed by the
single identity
\[
 r_f(u,v)=c(u,v)-f(u,v)+f(v,u),
\]
which holds for every ordered pair of distinct vertices,
whether or not it is an original arc of $N$.

A forward residual arc allows us to increase the flow on
an original arc; a reverse residual arc allows us to cancel
previously sent flow. Thus sending flow along a residual
$p$--$q$ path increases the value of $f$. This operation
is called \emph{augmentation}. The following lemma recalls the standard fact that routing a flow of value $\operatorname{val}(f)$ decreases the flow value in the residual graph by the same amount. (See, for example, {\cite[Lemma 26.4]{cormen1990introduction}}.)

\begin{lemma}
\label{lem:residual-cut-capacity}
Let $f$ be a feasible $p$--$q$ flow in a network $N$
with capacities $c(u,v)$ and residual capacities
$r_f(u,v)$. For every $S\subseteq V(N)$ with
$p\in S$ and $q\notin S$,
\[
 \sum_{\substack{u\in S\\v\notin S}}r_f(u,v)
 =
 \sum_{\substack{u\in S\\v\notin S}}c(u,v)
 -\operatorname{val}(f).
\]
Here capacities and flow values on absent arcs are zero.
\end{lemma}




\begin{definition}[Level graph]
Suppose that $q$ is reachable from $p$ in $N_f$.
Let $d(v)$ be the distance from $p$ to $v$ in $N_f$,
counting arcs, and set $h=d(q)$.
The level graph retains the arcs $u\to v$ satisfying
\[
 d(v)=d(u)+1\leq h,
\]
with their residual capacities.
\end{definition}

Every $p$--$q$ path in the level graph has length $h$.
Conversely, every shortest $p$--$q$ path in $N_f$ lies
in the level graph. The level graph therefore lets us
focus on all currently shortest paths at once.

\begin{definition}[Blocking flow]
A blocking flow is a feasible $p$--$q$ flow in the level
graph that saturates at least one arc on every $p$--$q$
path in that graph. An arc is saturated when the flow
through it equals its capacity in the level graph.
\end{definition}

For example, suppose the level graph consists of two
internally disjoint $p$--$q$ paths, all of whose arcs have
capacity one. Sending one unit along just one path is
not blocking, because the other path remains available.
Sending one unit along each path gives a blocking flow.

A blocking flow eliminates every currently shortest
residual path. Longer residual paths may remain, so
further augmentation may still be necessary.
\subsubsection{Implementing a blocking-flow phase}

We use the blocking-flow procedure of Anand, Saranurak,
and Wang~\cite[Claim~4.3]{ASW}. We first specify its input
and oracle access, and then explain how to implement
these oracles in our setting.

\paragraph{Input and oracle access.}
We state the blocking-flow subroutine for a general directed
network on $n+2$ vertices, matching the size of our flow
network $N$. For this statement, the capacities may be
arbitrary nonnegative integers, and the source and sink
are denoted by $p$ and $q$.
The current feasible integral flow $f$ is given explicitly:
we know every arc carrying positive flow and its flow
value; unlisted arcs carry zero flow.

The residual network is given implicitly through
the following two oracles:
\begin{enumerate}
\item \emph{Singleton Bipartite Independent Set (BIS) query:}
      given a vertex $u$ and a set
      $B\subseteq V(N)\setminus\{u\}$, determine whether
      $r_f(u,v)>0$ for some $v\in B$.
\item \emph{Residual-capacity query:}
      given an ordered pair $(u,v)$, return $r_f(u,v)$,
      with zero indicating that the residual arc is absent.
\end{enumerate}
Both oracles refer to the current residual network
and remain available as the algorithm augments $f$. The following lemma follows directly from the proof of \cite[Claim~4.3]{ASW}. Their claim allows arbitrary BIS
queries and assumes that the original flow network is undirected. However, their proof uses only the oracle operations specified above and does not rely on the
residual network arising from an undirected network. It therefore applies to directed flow networks as well.

\begin{lemma}[Adapted from {\cite[Claim~4.3]{ASW}}]
\label{lem:asw-blocking}
With the input and oracle access described above,
there is a deterministic algorithm that either
certifies that $f$ is maximum using $\wtO(n)$ oracle
calls, or computes an explicit integral blocking
flow $g$ and augments $f$ by $g$ using
\[
 \wtO\bigl(n+h\,\alpha\bigr)
\]
oracle calls in total, where $\alpha$ is the value of $g$. Here $h$ is the length of a
shortest residual $p$--$q$ path before augmentation.
The updated flow is maintained explicitly.
\end{lemma}




\paragraph{Implementing the oracles in our network.}
Return to the flow network $N$ defined above.
Its terminal arcs and their capacities are explicitly
known. Its internal arcs are the edges of $H$,
directed according to the orientation rule $\mathcal D$.
We maintain the current integral flow $f$ explicitly,
initially as the zero flow.

Consider an edge $\{u,v\}$ of $H$, directed from $u$
to $v$ by $\mathcal D$. Its capacity is one, so
$f(u,v)$ is either zero or one.
If $f(u,v)=0$, its only residual arc is $u\to v$,
of capacity one. If $f(u,v)=1$, its only residual
arc is $v\to u$, also of capacity one.
Thus the internal residual network is itself an
orientation of $H$.

Define the orientation rule $\mathcal D_f$ by reversing
$\mathcal D$ on every internal pair carrying one unit
of flow and retaining its direction on every other pair.
In particular, $\mathcal D_f$ agrees with $\mathcal D$
on non-edges, since these carry no flow.
The rule $\mathcal D_f$ can be evaluated without queries,
using $\mathcal D$ and the explicitly stored flow.
For distinct $u,v\in V(H)$, we therefore have
\[
 r_f(u,v)=
 \begin{cases}
  1, & \{u,v\}\in E(H)\text{ and }
       \mathcal D_f\text{ directs }u\to v,\\
  0, & \text{otherwise}.
 \end{cases}
\]

\begin{lemma}
\label{lem:residual-oracles}
Given the orientation rule $\mathcal D$ and the
explicit current integral flow $f$, each oracle call
in Lemma~\ref{lem:asw-blocking} can be implemented
using at most three undirected cut queries to $H$.
\end{lemma}

\begin{proof}
Let $Q(S)$ denote the undirected cut value of $S$ in $H$.

\emph{Residual-capacity queries.}
If either endpoint is a terminal, the original
capacities in both directions and their flow values
are explicitly known, so the residual capacity
requires no queries.

Now suppose $u,v\in V(H)$.
If $\mathcal D_f$ directs the pair from $v$ to $u$,
then $r_f(u,v)=0$.
Otherwise, $r_f(u,v)$ equals the indicator that
$\{u,v\}$ is an edge of $H$, namely
\[
 r_f(u,v)
 =\frac{Q(\{u\})+Q(\{v\})-Q(\{u,v\})}{2}.
\]
This requires at most three cut queries.

\emph{Singleton BIS queries.}
Given $u$ and $B$, first check the possible residual
arcs from $u$ into $B$ having at least one terminal
endpoint. Their capacities are explicitly known.
If any is positive, answer yes.
If $u$ is a terminal and none is positive, answer no,
since this check covers every possible residual arc
from $u$ into $B$, including reverse arcs.

It remains to consider $u\in V(H)$ and possible
neighbours in $B\cap V(H)$. Form the set
\[
 C=\{v\in B\cap V(H):
       \mathcal D_f\text{ directs }u\to v\}.
\]
The set $C$ is determined without queries.
By the definition of $\mathcal D_f$, a residual arc
from $u$ into $B\cap V(H)$ exists exactly when $H$
has an edge between $u$ and $C$.
The answer is therefore yes exactly when
\[
 \frac{Q(\{u\})+Q(C)-Q(\{u\}\cup C)}{2}>0.
\]
This requires at most three cut queries.
\end{proof}

Combining the two lemmas gives the subroutine we will
use in each phase of Dinic's algorithm.

\begin{lemma}[A blocking-flow phase in $N$]
\label{lem:blocking-phase}
Given the orientation rule $\mathcal D$ and an explicit
feasible integral flow $f$ in the repair network $N$,
a deterministic algorithm either certifies that $f$
is maximum using $\wtO(n)$ cut queries, or computes
an explicit integral blocking flow $g$ and augments
$f$ by $g$ using
\[
 \wtO(n+h\alpha)
\]
undirected cut queries to $H$, where $h$ is the current
residual $p$--$q$ distance and
$\alpha=\operatorname{val}(g)$.
The updated flow is maintained explicitly.
\end{lemma}

\begin{proof}
Apply Lemma~\ref{lem:asw-blocking}, implementing its
oracles by Lemma~\ref{lem:residual-oracles}.
After each augmentation, the updated explicit flow
determines the new rule $\mathcal D_f$ without
further queries.
\end{proof}
\subsubsection{Completing Dinic's algorithm}

We now use Lemma~\ref{lem:blocking-phase} to compute a
maximum flow in $N$. Recall that its maximum flow value
is $K=\Delta/4$. The analysis uses BFS layers, as in
\cite[Section~4.2]{ASW}.

\paragraph{The algorithm.}
Start with the zero flow.
While the residual network contains a $p$--$q$ path,
apply Lemma~\ref{lem:blocking-phase} to compute a
blocking flow and augment the current flow.
Maintain the flow explicitly between phases.
When no residual $p$--$q$ path remains, return
the current flow.

All flows remain integral. Whenever a residual $p$--$q$ path exists, the subroutine
finds a blocking flow of positive integral value.
When no such path exists, the current flow is maximum
and therefore has value $K$.
We use the following standard property of Dinic's
algorithm to bound the number of blocking-flow phases.

\begin{lemma}[\cite{D06}]
\label{lem:distance-increase}
After a blocking-flow phase, the shortest residual
$p$--$q$ distance strictly increases, unless $q$
becomes unreachable.
\end{lemma}



\paragraph{Remaining flow at large distance.}
The next lemma bounds the flow still to be sent once the residual distance becomes large. This closely follows the blocking-flow analysis of Even and Tarjan \cite{ET75} on unit capacity networks; we provide a proof for completeness.

\begin{lemma}[Adapted from {\cite[Lemma 3]{ET75}}; explicit in {\cite[Lemma B.3]{blumenstock2016fast}}]
\label{lem:remaining-flow-distance}
Let $f$ be the current flow, let
$R=K-\operatorname{val}(f)>0$, and let $h$ be the
shortest residual $p$--$q$ distance.
If $h\geq4$, then $
 R\leq \frac{4n^2}{h^2}.$
\end{lemma}
\begin{proof}
Let $L_j$ be the set of vertices at residual distance
$j$ from $p$. The layers $L_1,\ldots,L_{h-1}$ contain
only vertices of $H$. Thus
\[
 \sum_{j=1}^{h-2}
 \bigl(|L_j|+|L_{j+1}|\bigr)\leq 2n.
\]
For some $i\in\{1,\ldots,h-2\}$, therefore,
\[
 |L_i|+|L_{i+1}|
 \leq \frac{2n}{h-2}.
\]

Consider the residual cut with source side
$S=L_0\cup\cdots\cup L_i$.
Every residual arc leaving $S$ goes from $L_i$
to $L_{i+1}$: an arc cannot increase distance
from $p$ by more than one.
Neither of these layers contains a terminal.
Hence the large capacities of terminal arcs
do not contribute to this cut.

There is at most one edge of $H$ between any pair,
and its total residual capacity in both directions
is one. The residual capacity of the cut is
therefore at most
\[
 |L_i||L_{i+1}|
 \leq \frac{(|L_i|+|L_{i+1}|)^2}{4}
 \leq \frac{n^2}{(h-2)^2}
 \leq \frac{4n^2}{h^2}.
\]

By Lemma~\ref{lem:residual-cut-capacity}, the residual
capacity of this cut equals its original capacity
minus $\operatorname{val}(f)$.
Since $N$ admits a flow of value $K$, the original
cut capacity is at least $K$. Hence
\[
 R=K-\operatorname{val}(f)
 \leq \sum_{\substack{u\in S\\v\notin S}}r_f(u,v)
 \leq \frac{4n^2}{h^2},
\]
as required.
\end{proof}

\paragraph{The total query complexity.}
We now finish the proof of
Lemma~\ref{lem:repair-maxflow}(b).

\begin{proof}[Proof of Lemma~\ref{lem:repair-maxflow}(b)]
The algorithm returns an explicit integral flow of
value $K$, and hence a maximum flow.
For phase $i$, write $h_i$ for its initial residual
distance, $\alpha_i$ for the amount of flow it sends,
and $R_i$ for the amount remaining before the phase.
Thus
\[
 R_1=K,\qquad
 R_{i+1}=R_i-\alpha_i,
 \qquad
 \sum_i\alpha_i=K.
\]
Let $P$ count only the successful blocking-flow phases.
By Lemma~\ref{lem:blocking-phase}, the total query
complexity is
\[
 \wtO\left(n(P+1)+\sum_{i=1}^{P}h_i\alpha_i\right).
\]
The additional $n$ term accounts for the final check
that no residual $p$--$q$ path exists.
We bound $P$ and $\sum_i h_i\alpha_i$ separately.
If $K=0$, the algorithm terminates after the initial
$\wtO(n)$-query check. Henceforth assume $K>0$.

\emph{Number of phases.}
Set $d=\max\{4,\lceil n^{2/3}\rceil\}$.
By Lemma~\ref{lem:distance-increase}, there are at
most $d$ phases with $h_i<d$.
At the start of the first phase with $h_i\geq d$,
Lemma~\ref{lem:remaining-flow-distance} gives
\[
 R_i\leq \frac{4n^2}{d^2}.
\]
Every phase from this point onwards, including the
current phase, sends at least one unit. Hence there
are at most $4n^2/d^2$ such phases.
Consequently,
\[
 P\leq d+\frac{4n^2}{d^2}=O(n^{2/3}).
\]

\emph{The sum of path lengths.}
We first show that $K\leq |E(H)|\leq n^2$.
Recall that $D^*$ is an orientation of $H$ with
imbalance $\bb^*$. The orientation induced by
$\mathcal D$ and the orientation $D^*$ have the same
underlying graph $H$. Their imbalances therefore satisfy
\[
 \bb_D(v)=2\deg_{D}^+(v)-\deg_H(v),
 \qquad
 \bb^*(v)=2\deg_{D^*}^+(v)-\deg_H(v).
\]
Subtracting these identities gives
\[
 \ve(v)=\frac{\bb_D(v)-\bb^*(v)}{2}
       =\deg_{D}^+(v)-\deg_{D^*}^+(v).
\]
In particular, whenever $\ve(v)>0$, we have
$\ve(v)\leq\deg_{D}^+(v)$, since outdegrees are
nonnegative. Therefore,
\[
 K=\sum_{v:\ve(v)>0}\ve(v)
 \leq \sum_{v:\ve(v)>0}\deg_{D}^+(v)
 \leq \sum_{v\in V(H)}\deg_{D}^+(v)
 =|E(H)|.
\]
The last equality holds because each edge of $H$
is directed out of exactly one endpoint in $D$.
Finally, $H$ is simple, so $|E(H)|\leq\binom n2\leq n^2$.

For $h_i\geq4$,
Lemma~\ref{lem:remaining-flow-distance} implies
$h_i\leq 2n/\sqrt{R_i}$.
For $h_i<4$, the inequality $R_i\leq K\leq n^2$
instead gives $h_i\leq 3n/\sqrt{R_i}$.
Thus, in every phase,
\[
 h_i\leq \frac{3n}{\sqrt{R_i}}.
\]

Since $x^{-1/2}\geq R_i^{-1/2}$ for
$R_{i+1}<x\leq R_i$, we have
\[
 \frac{\alpha_i}{\sqrt{R_i}}
 \leq
 \int_{R_{i+1}}^{R_i}\frac{dx}{\sqrt{x}}.
\]
The intervals partition $(0,K]$, so
\[
 \sum_i h_i\alpha_i
 \leq 3n\sum_i\frac{\alpha_i}{\sqrt{R_i}}
 \leq 3n\int_0^K\frac{dx}{\sqrt{x}}
 =6n\sqrt K.
\]

Combining the two bounds, the total number of
undirected cut queries to $H$ is
\[
 \wtO(n^{5/3}+n\sqrt K)
 =
 \wtO(n^{5/3}+n\sqrt{\Delta}),
\]
as required.
\end{proof}

\subsection{Why explicit flow recovery depends on
\texorpdfstring{$K$}{K}}

We show that explicitly recovering a correction flow
can require $\Omega(n\sqrt K/\log n)$ cut queries.
The construction adapts the layered-graph lower bound
of Kenneth-Mordoch and Krauthgamer~
\cite[Lemma~1.8 and Appendix~B]{Kenneth-Mordoch26b}, whose bound
corresponds to $K=\Theta(n)$.

The idea is to construct many graphs with the same
orientation rule and target imbalance, but such that
every valid correction flow reveals all the edges
of the unknown graph.

\paragraph{A family of graphs.}
Fix integers $\ell\geq2$ and even $w\geq2$, and set
$n=\ell w$. Partition the vertices into $\ell$ layers
$L_1,\ldots,L_\ell$, each containing $w$ vertices.
Within each layer, fix a partition into $w/2$ pairs.
The layers and these pairings are known to the algorithm.

We place edges only between consecutive layers.
For each pair $\{a,b\}\subseteq L_i$ and each pair
$\{c,d\}\subseteq L_{i+1}$, choose independently
one of the following two sets of edges:
\[
 \{ac,bd\}
 \qquad\text{or}\qquad
 \{ad,bc\}.
\]
Thus each pair of pairs contributes one independent
binary choice. There are
\[
 M=(\ell-1)\left(\frac w2\right)^2
\]
such choices, giving a family of $2^M$ distinct
simple graphs $H$. The algorithm does not know
these choices; it accesses $H$ through cut queries.

Regardless of the choices, every vertex has exactly
one neighbour in each pair of the next layer, and hence
$w/2$ neighbours in that layer. Similarly, it has
$w/2$ neighbours in the previous layer whenever
that layer exists.

\paragraph{The orientation rule and target imbalance.}
Use the same orientation rule $\mathcal D$ for every
graph in the family: direct every pair in consecutive
layers from the earlier layer to the later layer,
and fix arbitrary directions for all other pairs.
The induced orientation of $H$ therefore directs
every edge from $L_i$ to $L_{i+1}$.

The degree calculation above gives
\[
 \bb_D(v)=
 \begin{cases}
  w/2,  & v\in L_1,\\
  -w/2, & v\in L_\ell,\\
  0,    & \text{otherwise}.
 \end{cases}
\]
Set $\bb^*=-\bb_D$. This target is feasible:
reversing every edge gives the required imbalance.
The repair demand vector is
\[
 \ve=\frac{\bb_D-\bb^*}{2}
     =\bb_D.
\]
Thus each first-layer vertex has supply $w/2$,
each last-layer vertex has demand $w/2$, and
all other vertices have demand zero. The total supply is
\[
 K=w\cdot\frac w2=\frac{w^2}{2}.
\]
In particular, $\mathcal D$, $\bb^*$, and $\ve$ are
identical for all $2^M$ graphs. Supplying them to the
algorithm reveals none of the binary choices.

\paragraph{Why the flow reveals every edge.}
Sending one unit along every edge directed by
$\mathcal D$ gives a feasible correction flow:
each first-layer vertex sends $w/2$ units,
each last-layer vertex receives $w/2$ units,
and every intermediate vertex receives and sends
$w/2$ units.

We now show that every feasible correction flow
must send one unit along every edge.
Fix $i\in\{1,\ldots,\ell-1\}$ and consider the set
\[
 S_i=L_1\cup\cdots\cup L_i.
\]
The total demand $\sum_{v\in S_i}\ve(v)$ is $K$,
so any correction flow of divergence $\ve$ must
send net flow $K$ out of $S_i$.
All arcs crossing this cut point from $L_i$
to $L_{i+1}$. There are exactly
\[
 w\cdot\frac w2=K
\]
such arcs, each of capacity one.
Consequently, every one of them must carry one
unit of flow.

This holds for every boundary between consecutive
layers, so every edge of $H$ carries one unit.
An explicit correction flow therefore lists every
edge of $H$. In particular, it reveals which of
the two matchings was chosen in each block,
and hence recovers all $M$ binary choices.

\paragraph{The query lower bound.}
An undirected cut query to a simple graph on $n$
vertices returns an integer between zero and $n^2$.
A deterministic algorithm making at most $Q$ queries
therefore has at most
\[
 (n^2+1)^Q
\]
possible answer transcripts, even when its queries
are adaptive.

For any fixed transcript, the algorithm produces
one output. Since a correct explicit flow determines
$H$ uniquely, this output can be correct for at most
one graph in the family.
Under the uniform distribution on the $2^M$ graphs,
the algorithm's success probability is therefore
at most
\[
 \frac{(n^2+1)^Q}{2^M}.
\]

The same bound holds for a randomised algorithm
with a worst-case budget of $Q$ queries: fix its
random choices, apply the preceding bound, and
then average over those choices.
Thus success probability at least $2/3$ on every
input requires
\[
 \frac{(n^2+1)^Q}{2^M}\geq\frac23,
\]
and hence
\[
 Q=\Omega\left(\frac{M}{\log n}\right).
\]
Finally, $\ell\geq2$, $n=\ell w$, and $K=w^2/2$ give
\[
 M=\frac{(\ell-1)w^2}{4}
 \geq \frac{\ell w^2}{8}
 =\frac{nw}{8}
 =\Omega(n\sqrt K).
\]
Therefore explicit correction-flow recovery requires
\[
 \Omega\left(\frac{n\sqrt K}{\log n}\right)
\]
cut queries for this family.

This lower bound concerns explicit flow recovery,
not the task of producing a corrected orientation rule.
Indeed, for this particular family, simply reversing
the given rule $\mathcal D$ already achieves the
target imbalance without any queries.

\input{general-digraphs}

\end{document}

%% file: header.tex
\usepackage[T1]{fontenc}
\usepackage[english]{babel}
\usepackage{verbatim}
\usepackage{bm}
\usepackage{amsmath,amstext,amsthm,amssymb}
\usepackage{float}
\usepackage{xspace}
\usepackage{graphicx}
\usepackage{mathtools}
\usepackage[font={small,it}]{caption}
\usepackage{subcaption,array}
\usepackage{wrapfig}
\usepackage{booktabs}
\usepackage{makecell}
\usepackage{array}
\usepackage{multirow}
\usepackage[usenames,dvipsnames]{xcolor}
\usepackage{boxedminipage,wrapfig}
\usepackage{pifont}
\usepackage[letterpaper,top=1in,bottom=2cm,left=1in,right=1in,marginparwidth=1.75cm,footskip=1cm]{geometry}
\usepackage{amsmath, amssymb}
\usepackage{amsthm}
\usepackage{comment}
\usepackage{mdframed}
\usepackage{xspace}
\usepackage{multicol}
\usepackage{tabularx}
\usepackage{makecell}

\usepackage[ruled,vlined,linesnumbered,algonl,procnumbered]{algorithm2e}
\SetEndCharOfAlgoLine{}
\SetKwComment{Comment}{\footnotesize$\triangleright$\ }{}

\SetCommentSty{mycommfont}

\usepackage{enumitem}
\usepackage{nicefrac}

\usepackage[colorlinks=true, allcolors=blue]{hyperref}
\usepackage{thmtools}
\usepackage{thm-restate}

\newtheorem{theorem}{Theorem}[section]
\newtheorem{lemma}[theorem]{Lemma}

\newtheorem{corollary}[theorem]{Corollary}

\newtheorem{observation}[theorem]{Observation}

\newtheorem{claim}[theorem]{Claim}
\newtheorem*{claim*}{Claim}
\newtheorem{definition}[theorem]{Definition}

\newtheorem{remark}[theorem]{Remark}

\usepackage{thmtools}

\allowdisplaybreaks

\declaretheoremstyle[
    bodyfont=\normalfont  
]{upright}

\usepackage{graphicx}

\newcommand{\eps}{\varepsilon}

\newcommand{\calB}{\mathcal{B}}

\newcommand{\calD}{\mathcal{D}}

\newcommand{\calP}{\mathcal{P}}

\DeclareMathOperator{\poly}{poly}

\newcommand{\bl}{\boldsymbol{\ell}}
\newcommand{\bv}{\mathbf{v}}
\newcommand{\bu}{\mathbf{u}}
\newcommand{\by}{\mathbf{y}}
\newcommand{\bz}{\mathbf{z}}
\newcommand{\bx}{\mathbf{x}}
\newcommand{\bh}{\mathbf{h}}

\DeclareMathOperator{\Reg}{Reg}

\usepackage[colorinlistoftodos,textsize=tiny,textwidth=2cm,color=green!50!gray]{todonotes}

\newlength\hwhatcwidth
\newlength\hwhatuwidth
\newcommand*\halfwidehat[1]{%
  \settowidth\hwhatcwidth{\ensuremath{#1}}%
  \settowidth\hwhatuwidth{\ensuremath{\mathrm{#1}}}%
  \addtolength\hwhatuwidth{-\hwhatcwidth}
  \makebox[0pt][l]{
    \kern\dimexpr0.25\hwhatcwidth-0.667\hwhatuwidth\relax
    \ensuremath{\widehat{\vphantom{#1}\rule{0.5\hwhatcwidth}{0pt}}}
  }#1
}

\usepackage{todonotes}

\newcommand{\eat}[1]{}

\usepackage[nameinlink]{cleveref}

\crefname{observation}{observation}{observations}
\Crefname{observation}{Observation}{Observations}
\Crefname{proposition}{Proposition}{Propositions}

\Crefname{algocf}{Algorithm}{Algorithms}
\crefname{algocfline}{line}{lines}
\Crefname{invariant}{Invariant}{Invariants}
\Crefname{claim}{Claim}{Claims}
\Crefname{subclaim}{Subclaim}{Subclaims}



%% file: abstract.tex
\begin{abstract}
We study $s$--$t$ reachability in an unweighted directed graph
on $n$ vertices, accessed through a directed cut oracle:
given $S\subseteq V$, the oracle returns the number of edges
directed from $S$ to $V\setminus S$. Directed cut functions
are submodular, and reachability asks whether the minimum
$s$--$t$ cut has positive value. This connects the problem
to the query complexity of submodular function
minimisation. Whether or not reachability can be solved in truly subquadratic query complexity has been an outstanding open question for a decade. 

We give a simple algorithmic blueprint with two randomised implementations.
The first uses $\widetilde O(n^{9/5})$ outgoing-cut queries;
the second improves this bound to $\widetilde O(n^{5/3})$.
The common idea is to construct an alternative orientation
with the same underlying undirected graph and vertex-wise degree
imbalances, thereby preserving reachability.
We first approximate the target imbalances using an averaging and rounding procedure, then correct the remaining error with a flow.

An earlier manuscript, made available in August 2026 in an online AI repository,
reports an $\widetilde O(n^{3/2})$ query bound.
Our contribution is a simple and comprehensible alternative approach that
already achieves a polynomial saving over quadratic
query complexity.
\end{abstract}

%% file: introduction.tex
\section{Introduction}\label{sec:problem}

\paragraph{Submodular function minimisation.} Submodular function minimisation (SFM) is a classical problem
in combinatorial optimisation. Given a submodular function
$f:2^V\to\mathbb R$ through a value oracle, the task is to find
a set minimising $f$ while making as few oracle queries as
possible. Since the seminal work of Gr\"otschel, Lov\'asz, and Schrijver \cite{GrotschelLS81} in 1981, who showed that SFM can be solved in a polynomial number of queries, numerous works have advanced our understanding of the exact complexity of SFM~\cite{Lovasz82,Cunningham85,GLS1988,Schrijver00,IwataFF01,FleischerI03,Iwata03,Vygen03,Harvey08,IwataO09,Orlin09,LeeSW15,ChakrabartyLS0W19,AxelrodLS20,ChakrabartyGJS22,Jiang23}. The current state of the art is an algorithm of complexity $\tilde O(n^2)$~\cite{Jiang21, EneN2026} whereas the strongest known
lower bound for randomised algorithms is only
$\Omega(n)$ \cite{Harvey08}.\footnote{For deterministic algorithms, an
$\Omega(n\log n)$ lower bound is known~\cite{ChakrabartyGJS22}.}
Closing this polynomial gap remains elusive and is an outstanding open problem (as mentioned in many of these papers). Understanding this gap motivates studying structured classes
of submodular functions, both as candidates for stronger lower
bounds and as sources of ideas for better algorithms.

\paragraph{Graph cut queries.} Graph cut functions provide a natural setting for this
investigation. A value query to a cut function specifies a
vertex set and returns the number, or total weight, of edges
crossing the corresponding cut. This gives the \emph{cut-query
model}, introduced by Rubinstein, Schramm, and
Weinberg~\cite{RubinsteinSW18}, in which an algorithm accesses the graph
only through such queries. The goal is to exploit the structure
of graph cuts without assuming direct access to individual
edges. For undirected graphs, this has led to efficient
algorithms: connectivity, the global minimum cut, and bipartite matching
can be computed using $\wtO(n)$ randomised cut
queries~\cite{ApersEGLMN22, MukhopadhyayN20, blikstadBEMN22}; $s,t$-minimum cut can be computed using $\wtO(n^{5/3})$ deterministic cut queries \cite{ASW} and using $\wtO(n^{8/5})$ randomised cut queries \cite{JiangNS26}. The latter was quickly improved to an all-pairs minimum cut algorithm using $\wtO(n^{3/2})$ randomised cut queries \cite{Kenneth-Mordoch26b}. 

Directed graphs present a different challenge. Consider the basic
question of reachability, i.e., whether $t$ is reachable from $s$. For undirected graphs, this is a question of connectivity which is easy in the cut-query
model: a standard BFS or DFS can be implemented using
$O(n\log n)$ cut queries. For directed graphs, however, the
same approach does not work. Although outgoing-cut queries
to a directed graph $G$ allow us to simulate undirected cut
queries to its underlying undirected graph $H$, they do not
reveal the directions of individual edges. In fact, even
knowing the answers to all outgoing-cut queries does not
suffice to recover these directions. For example, reversing
all edges of a directed triangle leaves every outgoing-cut
value unchanged. Thus, the straightforward graph-search
approach for undirected connectivity does not extend to
directed reachability.

Nevertheless, directed reachability is closely related to
submodular function minimisation. A vertex $t$ is unreachable
from $s$ precisely when there exists a set $S$ containing
$s$ but not $t$ with no outgoing edge. Thus deciding
reachability is equivalent to determining whether there is a cut separating $s$ and $t$ that has value $0$.
Fixing $s$ inside the set and $t$ outside gives a
submodular minimisation problem on the remaining $n-2$
vertices. Consequently, a superlinear randomised cut-query
lower bound for directed reachability would also imply
a superlinear value-query lower bound for general submodular
function minimisation.


Recent work has obtained efficient reachability algorithms
for directed acyclic graphs~\cite{KhannaPS26,BalsTN26}. These works
highlight the general directed case, for which the bounds
they record are $\Omega(n)$ and $O(n^2/\log n)$.
The question is therefore whether one can obtain a polynomial,
rather than merely logarithmic, saving over quadratic query
complexity. Here we develop a simple algorithmic blueprint
that achieves such a saving for general unweighted directed graphs. 


\subsection{Our contribution}
Our main contributions are two directed cut query algorithms for $s$--$t$ reachability on unweighted digraphs, one achieving $\wtO(n^{9/5})$ query complexity, thus already breaking the quadratic barrier, and an improvement of this algorithm to achieve $\wtO(n^{5/3})$ query complexity. To state these results here, we assume that the directed graph is oriented (i.e., no 2-cycle is present). This constraint can be removed with minimal modifications to the algorithms, as described in Appendix \ref{app:general-digraphs}.

\begin{theorem}[Warm-up protocol]\label{thm:warmup}
There is a randomised zero-error algorithm for directed
$s$--$t$ reachability in oriented graphs using
$\wtO(n^{9/5})$ outgoing-cut queries in expectation and with
high probability.
\end{theorem}

\begin{theorem}[Faster protocol]\label{thm:main}
There is a randomised zero-error  algorithm for directed
$s$--$t$ reachability in oriented graphs using
$\wtO(n^{5/3})$ outgoing-cut queries in expectation and with
high probability.
\end{theorem}

The common idea is to construct an \emph{orientation rule} that assigns a direction to every unordered pair of distinct
    vertices, and whose
induced orientation of the underlying undirected graph of $G$ has the same
vertex imbalances as $G$. This preserves all outgoing-cut
values and hence reachability (as observed in \cite{GhoshKMM26}). We first construct a sequence of (possibly fractional) orientation
rules whose average induces vertex imbalances close to those of
$G$. We then round this average to obtain an integral orientation
rule and repair the remaining imbalance by reversing edges carrying
a suitable flow. The two implementations differ in how they
construct and round the average; the reduction and flow repair
are the same.

\begin{remark}[Why two implementations?]
We include the warm-up for its pedagogical value: it presents
the main ideas in a simple implementation that already achieves
subquadratic query complexity. This lets the reader understand
the blueprint before encountering the additional machinery
needed for the faster bound. The faster implementation then
shows how to improve the bound within the same blueprint,
by refining the averaging and rounding steps while leaving
the reduction and flow repair unchanged.
\end{remark}

\begin{remark}[Recent work] A recent unpublished AI-generated manuscript, which appeared
in an online repository in August 2026, reports an
$\wtO(n^{3/2})$ algorithm~\cite{CutReachability2026}.
However, its technical exposition is difficult to follow,
making the underlying algorithmic ideas and their analysis
hard to understand and verify. Our algorithms were also initially discovered through interactive prompting of LLMs (see \Cref{sec:ai} for our AI disclosure), with the explicit goal of finding a
substantially simpler and more transparent approach that
already achieves a subquadratic upper bound, even though our bound is weaker.  We sketch the
main technical differences from~\cite{CutReachability2026}
in \Cref{sec:blueprint}.
\end{remark}
\paragraph{Organisation.} We first mention the cut query primitives that we use in this work in \Cref{sec:prelims}. These primitives are standard, and readers who are familiar can jump directly to \Cref{sec:blueprint}, which presents the blueprint and its
warm-up implementation. \Cref{sec:warmup} proves the
averaging and rounding guarantees that are used but not proved in \Cref{sec:blueprint}.
Appendices~\ref{sec:improved} and \ref{app:balanced-completion} give the faster implementation. We provide the adaptation of \cite{ASW} that we use in our work in Appendix \ref{app:flow}.

%% file: blueprint.tex
\input{blueprint-modification}

\subsection{Constructing an approximate orientation rule}
\label{sec:meta}

In this section, we discuss Step \ref{bp:approx-imbalance} of the blueprint. Recall from \Cref{defn:implicit-orient} that the input is a simple undirected graph $H=(V,E)$ and a feasible
target imbalance vector $\bb^*$. Before constructing an orientation
rule matching $\bb^*$ exactly, we first find one whose induced
orientation approximates $\bb^*$. Both implementations use a
parameter $T$ specifying the number of averaging iterations.
Write $\varepsilon(T)$ for the total error bound guaranteed by
the chosen implementation: its output is an orientation rule
$\mathcal D$ whose induced orientation $D$ satisfies
\[
 \norm{\bb_D-\bb^*}_1\leq\varepsilon(T).
\]
We establish the bound $\varepsilon(T)$ in the analysis of each
implementation. This error determines the cost of the subsequent
repair step.

The two implementations use averaging and rounding, but differ
in the objects they compute. We introduce fractional orientation
rules to describe their error analysis.

\begin{definition}[Fractional orientation rule] \label{def:frac-orient-rule}
Fix the reference direction $u\to v$ whenever $u<v$. A
\emph{fractional orientation rule} $\bz$ assigns a value
$\bz(u,v)\in[-1,1]$ to every unordered pair $\{u,v\}$, written
with $u<v$. The values $+1$ and $-1$ encode the directions
$u\to v$ and $v\to u$, respectively. For an intermediate value,
$(1+\bz(u,v))/2$ is the weight assigned to $u\to v$, with the
remaining weight $(1-\bz(u,v))/2$ assigned to $v\to u$. The imbalance it induces
on $H$ is
\begin{equation}\label{eq:fractional-imbalance}
 \bb_{\bz}(v)=\sum_{w>v}c_{vw}\bz(v,w)
             -\sum_{u<v}c_{uv}\bz(u,v).
\end{equation}
\end{definition}
Since $\bb^*$ is feasible, there is an orientation $D^*$ of
$H$ with imbalance $\bb^*$. Encoding its directions by
$+1$ and $-1$, and assigning arbitrary values to absent pairs,
gives a fractional orientation rule $\bz^*$ with
$\bb_{\bz^*}=\bb^*$.

\subsubsection{Warm-up implementation using cut queries}
\label{sec:approx-cut-implementation}
\swagato{The algorithm has two parts: FTRL and rounding. The FTRL part is included in Algorithm 1. The rounding part is written separately, after a gap, and without any environment. This probably isn't ideal..?}
We now give the warm-up construction using undirected cut
queries to $H$. We state the two lemmas needed for its analysis (\Cref{lem:simple-average} and \Cref{lem:analysis-part2})
in this section and defer their proofs to \Cref{sec:warmup}.

\paragraph{Averaging using FTRL.}
Our goal is to generate a sequence of orientations whose
average imbalance is close to $\bb^*$; an individual orientation
in the sequence need not have small error.

The algorithm maintains a value $\by_t(v)$ at each vertex,
recording its accumulated imbalance error:
\[
 \by_t(v)
 =\eta\sum_{\tau<t}
   \bigl(\bb_{\bz_\tau}(v)-\bb^*(v)\bigr).
\]
Fix such a vertex $v$. If $\by_t(v)>0$, the previous orientations
have, on average, given $v$ an imbalance larger than its
target $\bb^*(v)$. To compensate for this excess, we would
like subsequent orientations to direct more edges into $v$,
since incoming edges reduce its imbalance. Conversely, if
$\by_t(v)<0$, its imbalance has been too small on average,
so we would like subsequent orientations to give it more
outgoing edges.

The direction of an edge must, however, take both endpoints
into account. Consider a pair $\{u,v\}$ with
$\by_t(u)<\by_t(v)$. The accumulated signed imbalance error
at $v$ is larger than at $u$, so we direct the pair from
$u$ to $v$. If this pair is an edge of $H$, it contributes
$-1$ to the imbalance at $v$ and $+1$ to the imbalance at $u$.
This explains the rule of directing each pair from the
smaller value to the larger one: it encourages later
orientations to compensate for the errors accumulated
in earlier iterations.

This choice of directions is important for the analysis, which is done using the follow-the-regularised-leader (FTRL) rule from online
optimisation~\cite{SS12,Trevisan19}. We give a brief overview here; more details will follow in \Cref{sec:warmup}.
Note that the loss vector used in \Cref{alg:simple-average} (Step \ref{step:loss})  is
$\bl_t=\bb^*-\bb_{\bz_t}$. Directing each pair from the
smaller $\by_t$-value to the larger one minimises
$\langle\by_t,\bb_{\bz_t}\rangle$ over all orientation rules.
Since $\bb^*$ is the imbalance of some orientation of $H$,
we have $\langle\by_t,\bb_{\bz_t}\rangle
\leq \langle\by_t,\bb^*\rangle$,
or equivalently,
$\langle\by_t,\bl_t\rangle\geq 0$.
Thus the loss incurred by the algorithm is nonnegative
in every iteration.

At this point, define the average fractional rule
\[
 \bar{\bz}:=\frac1T\sum_{t=1}^T\bz_t.
\]
We are to show that the average imbalance of the generated
orientations $\bb_{\bar{\bz}}$ is close to $\bb^*$.
Suppose otherwise. Consider a fixed vector
$\mathbf u=\operatorname{sign}(\bb_{\bar{\bz}} - \bb^*)$,
whose coordinates are the signs of this average error.
It can be shown that its cumulative loss, over all $T$ iterations, would be $-T\norm{\bb_{\bar{\bz}} - \bb^*}_1$. On the other hand, using a standard telescoping argument one can show that the
algorithm's cumulative loss cannot be much larger than
that of any fixed vector, in particular, that of $\mathbf u$. This, in addition to the fact that the algorithm's loss
is nonnegative, forces the average imbalance error
to be small.

\begin{restatable}[Simple averaging]{lemma}{simpleaveraging}
\label{lem:simple-average}
For every integer $1\leq T\leq n$, \Cref{alg:simple-average}
deterministically uses $O(nT)$ undirected cut queries and
generates orientation rules $\bz_1,\ldots,\bz_T$ whose
average $\bar{\bz}$ satisfies
\[
 \norm{\bb_{\bar{\bz}}-\bb^*}_1
 \leq \frac{4n^2}{\sqrt T}.
\]
\end{restatable}


%

\begin{algorithm}[t]
\caption{Averaging integral orientation rules}
\label{alg:simple-average}

\textbf{Input:} Undirected cut-query access to a simple graph
$H$ on $n$ vertices, a feasible target imbalance vector $\bb^*$,
and an integer $1\leq T\leq n$.

\textbf{Output:} Integral orientation rules $\bz_1,\ldots,\bz_T$.

\medskip
Set $\eta=\frac{1}{2n\sqrt{2T}}$,
and initialise $\by_1=\mathbf 0$. For each iteration
$t=1,\ldots,T$, do the following.
\begin{enumerate}
    \item Define the orientation rule $\bz_t$ by setting,
    for every $u<v$,
    \[
     \bz_t(u,v)=
     \begin{cases}
       +1, & \text{if }\by_t(u)\leq\by_t(v),\\
       -1, & \text{otherwise.}
     \end{cases}
    \]
    Thus each pair is directed from the vertex with the
    smaller value to the vertex with the larger value,
    breaking ties by directing from the smaller label
    to the larger one.

    \item \label{step:loss} Compute the imbalance vector $\bb_{\bz_t}$ of
    the induced orientation using $O(n)$ queries by
    \Cref{obs:imbalance}, and define the loss vector
    \[
     \bl_t:=\bb^*-\bb_{\bz_t}.
    \]

    \item \label{step:update}
    Update the vertex values using the FTRL rule:
    \[
     \by_{t+1}
       =\by_t-\eta\bl_t
       =\by_t+\eta(\bb_{\bz_t}-\bb^*).
    \]
\end{enumerate}
Return $\bz_1,\ldots,\bz_T$.
\end{algorithm}

Only the imbalance computation requires queries to $H$.

\paragraph{Independent rounding.}
For each unordered pair $\{u,v\}$ with $u<v$, independently
choose an iteration $t_{uv}$ uniformly from $\{1,\ldots,T\}$
and give the pair its direction in $\bz_{t_{uv}}$.
These choices define an orientation rule $\mathcal D$;
let $D$ be its induced orientation of $H$.

This procedure has exactly the same distribution as
independent rounding of $\bar{\bz}$. Indeed, for every $u<v$,
\[
 \Pr[(u,v)\in\mathcal D]
 =\frac{|\{t:\bz_t(u,v)=+1\}|}{T}
 =\frac{1+\bar{\bz}(u,v)}{2},
\]
and the choices for distinct pairs are independent.
Thus the algorithm samples according to $\bar{\bz}$
without explicitly forming it or evaluating its fractional
imbalance. The following lemma bounds the rounding error.

\begin{restatable}[Independent rounding]{lemma}{independentrounding}
\label{lem:analysis-part2}
For every integer $1\leq T\leq n$, the sampled orientation
rule induces an orientation $D$ satisfying
\[
 \norm{\bb_D-\bb_{\bar{\bz}}}_1
 \leq \frac{4n^2}{\sqrt T}
\]
with probability at least $1/2$.
\end{restatable}

\paragraph{Checking the total error.}
By \Cref{lem:simple-average,lem:analysis-part2}, with
probability at least $1/2$,
\[
 \norm{\bb_D-\bb^*}_1
 \leq \norm{\bb_D-\bb_{\bar{\bz}}}_1
      +\norm{\bb_{\bar{\bz}}-\bb^*}_1
 \leq \frac{8n^2}{\sqrt T}.
\]
We therefore set $\varepsilon(T):=\frac{8n^2}{\sqrt T}$.
After each rounding trial, compute $\bb_D$ using $O(n)$
queries by \Cref{obs:imbalance}. If
$\norm{\bb_D-\bb^*}_1\leq\varepsilon(T)$, return
$\mathcal D$; otherwise, repeat the rounding with fresh
randomness, keeping the same sequence $\bz_1,\ldots,\bz_T$.
This check uses only $\bb_D$ and $\bb^*$, not the fractional
imbalance $\bb_{\bar{\bz}}$.

Each trial succeeds with probability at least $1/2$, so
there are at most two trials in expectation and $O(\log n)$
trials with high probability. Sampling the directions
requires no queries, and checking each trial \swagato{`checking each trial'? also, the query complexity/cut query implementation is written in a separate titled section for other things, but I'm not sure whether we need to do that everywhere.} costs $O(n)$
queries. Including the averaging iterations, the total
cost is $O(nT)$ in expectation and $O(nT+n\log n)$ with
high probability, hence $\wtO(nT)$ in both cases.

Whenever the procedure returns, the induced orientation
satisfies $\norm{\bb_D-\bb^*}_1\leq\varepsilon(T)$.
We then proceed to the exact repair step.

\subsection{Repairing the imbalance by a flow}\label{sec:repair}

\Cref{sec:meta} constructs an orientation rule whose
induced orientation $D$ has imbalance close to $\bb^*$. We now
correct the remaining error while retaining an implicit
representation (Step \ref{bp:flow-repair} of the blueprint).

Reversing an edge $u\to v$ decreases the imbalance at $u$ by two
and increases the imbalance at $v$ by two. This suggests using a
flow to determine which edges to reverse. The following lemma
formalises this idea.
\begin{lemma}[Exact repair]\label{lem:abstract-repair}
Let $D$ be any orientation of $H$, with imbalance $\bb_D$.
Give each directed edge of $D$ unit capacity and set
$\ve=(\bb_D-\bb^*)/2$. Then:
\begin{enumerate}
    \item There exists an integral flow $\vf$ in $D$ whose
    net outgoing flow at every vertex $v$ equals $\ve(v)$.
    \item Reversing the edges carrying one unit of any such flow
    gives an orientation with imbalance $\bb^*$.
    \item The total supply is
    \begin{equation}\label{eq:repair-value}
    K=\sum_{v:\ve(v)>0}\ve(v)
     =\frac12\norm{\ve}_1
     =\frac14\norm{\bb_D-\bb^*}_1.
    \end{equation}
\end{enumerate}
\end{lemma}
\begin{proof}
Fix an orientation $D^*$ of $H$ with imbalance $\bb^*$,
which exists by feasibility. Since the total degree at each
vertex is fixed,
\[
\ve(v)=\frac{\bb_D(v)-\bb^*(v)}{2}
      =\deg_D^+(v)-\deg_{D^*}^+(v).
\]
Thus $\ve$ is integral. Now send one unit of flow along every edge of $D$ whose
direction differs from $D^*$. This respects the unit capacities.
For every vertex $v$, the net outgoing flow is exactly
$\deg_D^+(v)-\deg_{D^*}^+(v)=\ve(v)$, since edges whose
directions agree contribute nothing to either difference.
This proves the first claim.

Now consider any integral flow $\vf$ satisfying the first claim,
and let $D'$ be obtained by reversing its edges carrying one
unit. Each reversed outgoing edge decreases the imbalance by
two, while each reversed incoming edge increases it by two.
Consequently,
\[
\bb_{D'}(v)=\bb_D(v)-2\ve(v)=\bb^*(v),
\]
proving the second claim.

Finally, $\sum_v\ve(v)=0$, so the total supply equals
$\norm{\ve}_1/2=\norm{\bb_D-\bb^*}_1/4$.
\end{proof}

\begin{remark}[Flow conservation]
The flow in \Cref{lem:abstract-repair} need not satisfy the usual
flow conservation condition at every vertex, since the net
outgoing flow at $v$ is $\ve(v)$.
We can formulate this as a standard $s$-$t$ (or, in this case, $p$-$q$) flow problem by
introducing a super-source $p$ and a super-sink $q$. For every
vertex $v$ with $\ve(v)>0$, add an edge $p\to v$ of capacity
$\ve(v)$, and for every vertex with $\ve(v)<0$, add an edge
$v\to q$ of capacity $-\ve(v)$.
An integral flow in $D$ with the required net outgoing flow
at each vertex corresponds to a $p$-$q$ flow of value
$K=\sum_{v:\ve(v)>0}\ve(v)$. In this augmented network,
the usual flow conservation holds at every vertex other
than $p$ and $q$.
\end{remark}

\subsubsection{Implementation using cut queries}
\label{sec:flow-repair-implementation}

By \Cref{obs:imbalance}, we can compute $\bb_D$ using $O(n)$
undirected cut queries to $H$. This also determines the demand
vector $\ve=(\bb_D-\bb^*)/2$.
It remains to find the correction flow without learning all
edges of $H$. The flow subroutine in Appendix~\ref{app:flow}
adapts the blocking-flow implementation of Anand, Saranurak,
and Wang~\cite{ASW}, combining it with the Even-Tarjan analysis of blocking-flow on uncapacitated networks~\cite{ET75}. Together with \Cref{lem:abstract-repair},
it gives the following guarantee. 

\begin{lemma}[Cut-query repair]\label{lem:repair}
Given undirected cut-query access to $H$, a feasible target
imbalance vector $\bb^*$, and an orientation rule $\mathcal D$
on $V(H)$ inducing an orientation $D$, we can deterministically
construct an orientation rule whose induced orientation has
imbalance exactly $\bb^*$ using
\[
 \wtO\left(n^{5/3}+n\sqrt{\norm{\bb_D-\bb^*}_1}\right)
\]
undirected cut queries.
\end{lemma}
Below we provide a short proof sketch assuming an adaptation of~\cite{ASW}. The full proof is deferred to Appendix \ref{app:flow} (\Cref{lem:flowmain}).
\begin{proof}[Proof sketch]
Compute $\bb_D$ using $O(n)$ queries by \Cref{obs:imbalance},
and set $\ve=(\bb_D-\bb^*)/2$. Give each edge of $D$ unit
capacity. By \Cref{lem:abstract-repair}, there exists an
integral flow respecting these capacities whose net outgoing
flow at each vertex $v$ equals $\ve(v)$. The total supply is
\[
 K=\frac14\norm{\bb_D-\bb^*}_1.
\]
If $K=0$, return $\mathcal D$. Otherwise, the subroutine in
Appendix~\ref{app:flow}, adapted from~\cite{ASW}, computes
such a flow $\vf$ explicitly using
$\wtO(n^{5/3}+n\sqrt K)$ undirected cut queries to $H$.

Define a new orientation rule $\mathcal D'$ by reversing
the direction assigned by $\mathcal D$ to each edge carrying
one unit of flow, and leaving all other pairs unchanged.
Since $\vf$ is explicit, this requires no additional queries.
By \Cref{lem:abstract-repair}, the orientation induced by
$\mathcal D'$ on $H$ has imbalance $\bb^*$.

Substituting the value of $K$ and including the $O(n)$
queries used to compute $\bb_D$ gives the stated bound.
\end{proof}

In particular, if the orientation rule constructed in
\Cref{sec:meta} induces an orientation $D$ satisfying
$\norm{\bb_D-\bb^*}_1\leq\varepsilon(T)$, then
\Cref{lem:repair} gives a repair cost of
\[
 \wtO\bigl(n^{5/3}+n\sqrt{\varepsilon(T)}\bigr)
\]
additional undirected cut queries. Thus a smaller approximation
error reduces the error-dependent term in the repair bound.

\subsection{The resulting query bounds}

Fix the number of averaging iterations $T$. Let $Q_{\mathrm{av}}$
and $Q_{\mathrm{rd}}$ denote the undirected cut-query costs of
averaging and rounding, and write $\varepsilon=\varepsilon(T)$
for the resulting imbalance error bound. Combining
\Cref{lem:repair,lem:directed-to-undirected} gives a total
outgoing-cut query bound of
\begin{equation}\label{eq:framework-cost}
 O\bigl(Q_{\mathrm{av}}+Q_{\mathrm{rd}}\bigr)
 +\wtO\bigl(n^{5/3}+n\sqrt{\varepsilon}\bigr).
\end{equation}
The $O(n\log n)$ overhead of the reduction is absorbed by
the $n^{5/3}$ term.

The two implementations give the following guarantees.  The first two rows apply for every
integer $1\leq T\leq n$. The last two rows give the chosen value
of $T$ and the resulting query bound from~\eqref{eq:framework-cost}. All query bounds hold in expectation and with high probability.

\begin{center}
\small
\renewcommand{\arraystretch}{1.2}
\begin{tabular}{@{}lll@{}}
\toprule
 & Warm-up & Faster implementation\\
\midrule
$Q_{\mathrm{av}}(T)+Q_{\mathrm{rd}}(T)$
 & $\wtO(nT)$ & $\wtO(nT+n^{5/3})$\\
Imbalance error bound $\eps(T)$
 & $O(n^2/\sqrt T)$ & $O(n^2/T)$\\
Choice of $T$
 & $\lceil n^{4/5} \rceil$ & $\lceil n^{2/3}\rceil$\\
Total cut queries
 & $\wtO(n^{9/5})$ & $\wtO(n^{5/3})$\\
\bottomrule
\end{tabular}
\end{center}



%% file: blueprint-modification.tex
\section{Preliminaries} \label{sec:prelims}

We collect the definitions and cut-query operations used in the
rest of this section. Many of these are mentioned in previous works (for example, in \cite{RubinsteinSW18, MukhopadhyayN20, ApersEGLMN22, Kenneth-Mordoch26, JiangNS26, Kenneth-Mordoch26b, KhannaPS26}, to name a few) and are standard. We provide them here for completeness.

\paragraph{Graph notation.}
All graphs have labelled vertices and no loops. Throughout the
main text, directed graphs have at most one edge between each
unordered pair of vertices. For a directed graph $G$, its
undirected \emph{skeleton} is the simple undirected graph obtained
by removing the edge directions. The \emph{imbalance} at a vertex
$v$ is
\[
 \bb_G(v):=\deg_G^+(v)-\deg_G^-(v).
\]
For $S\subseteq V(G)$, its outgoing-cut value is
\[
 F_G(S)=|\{(u,v)\in E(G):u\in S,\ v\notin S\}|.
\]
An outgoing-cut query specifies $S$ and returns $F_G(S)$.

For a simple undirected graph $H=(V,E)$, write
\[
 C_H(S)=|\{\{u,v\}\in E:u\in S,\ v\notin S\}|.
\]
An undirected cut query specifies $S\subseteq V$ and returns
$C_H(S)$. For distinct vertices $u,v$, let $c_{uv}\in\{0,1\}$
indicate whether $\{u,v\}$ is an edge of $H$. Write $E_H[S]$ for the edges with
both endpoints in $S$, and $\delta_H(S)$ for the edges crossing the cut $S$. 

\begin{definition}[Orientation rule]\label{def:orientation-rule}
An \emph{orientation rule} $\mathcal D$ on a vertex set $V$
assigns a direction to every unordered pair of distinct vertices.
We write $(u,v)\in\mathcal D$ when the assigned direction is
$u\to v$. The direction of a pair can be evaluated from the data
specifying the rule.

For a simple undirected graph $H$ on $V$, the \emph{induced
orientation} $D$ is obtained by restricting $\mathcal D$ to the
edges of $H$:
\[
 E(D) =\{(u,v)\in\mathcal D:\{u,v\}\in E(H)\}.
\]
Here $D$ is a directed graph and $E(D)$ represents its edge set.
We say that $\mathcal D$ \emph{matches} an imbalance vector $\bb$
on $H$ if $\bb_D=\bb$.
\end{definition}

An imbalance vector $\bb$ is \emph{feasible} for $H$ if some
orientation of $H$ has imbalance vector $\bb$. The rule
$\mathcal D$ itself is defined on $V$, independently of which
pairs are edges of $H$.

In the following primitives, $H$ has $n$ vertices and is accessible
through undirected cut queries. Any supplied orientation rule or
predicate can be evaluated without queries to $H$.

\paragraph{Counting edges between sets.}
For disjoint sets $A,B\subseteq V$, define
\[
 c(A,B)=\sum_{u\in A,v\in B}c_{uv}.
\]
\begin{observation}[Edge-counting identity]\label{obs:edge-counting}
For any two disjoint sets $A,B\subseteq V$,
\begin{equation}\label{eq:rectangle}
 c(A,B)=\frac{C_H(A)+C_H(B)-C_H(A\cup B)}{2}.
\end{equation}
Consequently, $c(A,B)$ can be computed using three
undirected cut queries.
\end{observation}
Indeed, edges between $A$ and $B$ are counted twice in the
numerator, while the contributions of all other edges cancel.
In particular, for a vertex $v$ and a set
$X\subseteq V\setminus\{v\}$, we can count the neighbours of $v$
in $X$ using $O(1)$ queries. Taking $X$ to be a singleton also
determines whether a specified pair is an edge.

\paragraph{Finding a neighbour.}
The edge-counting identity also allows us to find a neighbour
within a specified set.

\begin{lemma}[Neighbour search]\label{lem:cut-neighbour}
Given a vertex $v$ and a set $X\subseteq V\setminus\{v\}$,
we can find a neighbour of $v$ in $X$, or certify that none
exists, using $O(\log n)$ undirected cut queries.
\end{lemma}
\begin{proof}
Use \Cref{obs:edge-counting} to compute
$c(\{v\},X)$. If this is zero, no such neighbour exists.
Otherwise, partition $X$ into two nearly equal sets $X_1,X_2$
and compute $c(\{v\},X_1)$. If this count is positive, continue
with $X_1$; otherwise, continue with $X_2$, which must contain
a neighbour. Each iteration uses $O(1)$ queries and reduces
the candidate set by approximately half. After $O(\log n)$
iterations, the remaining vertex is a neighbour of $v$.
\end{proof}

\paragraph{Subgraphs specified by known pairs.}
We will also need to work with subgraphs of $H$ without
learning all their edges. Suppose we are explicitly given a family
$\mathcal P\subseteq\binom{V}{2}$ of potential pairs. Let $J$ be the subgraph of $H$ consisting of the
edges belonging to $\mathcal P$, i.e.,
\[
 E(J)=E(H)\cap\mathcal P.
\]
The following lemma shows that several basic operations on
$J$ can be implemented using cut queries to $H$.

\begin{lemma}[Operations on subgraphs]\label{lem:subgraph-operations}
Let $J$ be defined as above. Using undirected cut queries
to $H$, we can:
\begin{enumerate}
    \item For any vertex $v$ and set
    $X\subseteq V\setminus\{v\}$, count the neighbours
    of $v$ in $X$ in $J$ using $O(1)$ queries, or find
    one such neighbour (or certify its absence) using
    $O(\log n)$ queries.

    \item Count the number of edges of $J$ using
    $O(n)$ queries.

    \item Enumerate all $a$ edges of $J$ using
    $\wtO(n+a)$ queries.

    \item Find a spanning forest of $J$ using
    $O(n\log n)$ queries.
\end{enumerate}
\end{lemma}

\begin{proof}
For a vertex $v$ and a set $X\subseteq V\setminus\{v\}$,
define
\[
 X_v=\{w\in X:\{v,w\}\in\mathcal P\}.
\]
The neighbours of $v$ in $X$ in $J$ are exactly the
neighbours of $v$ in $X_v$ in $H$. Thus, their number
is $c(\{v\},X_v)$, which can be computed using $O(1)$
queries by \Cref{obs:edge-counting}. Similarly,
\Cref{lem:cut-neighbour} finds one such neighbour,
or certifies its absence, using $O(\log n)$ queries.
This proves the first claim.

For the second claim, apply the first claim with
$X=V\setminus\{v\}$ to compute the degree of every
vertex in $J$. Summing the degrees and dividing by
two gives the number of edges of $J$.

To enumerate the edges of $J$, consider each vertex $v$
and repeatedly find and remove its neighbours $w>v$
from the candidate set until none remains. Each edge
is found exactly once. There are $a$ successful
searches and at most $n$ unsuccessful searches,
giving $\wtO(n+a)$ queries.

Finally, we can find a spanning forest of $J$ by
graph search. Starting from an arbitrary vertex,
repeatedly find neighbours outside the set of reached
vertices, marking a vertex finished when no such
neighbour exists. Once a component is exhausted,
restart from an unreached vertex. There are at most
$n-1$ successful searches and $n$ unsuccessful
searches, giving $O(n\log n)$ queries.
\end{proof}

Notice that the lemma does not imply that an arbitrary
cut query to $J$ can be simulated using $O(1)$ cut
queries to $H$.

\paragraph{Evaluating the imbalance of an induced orientation.}
\begin{observation}[Computing the imbalance vector]
\label{obs:imbalance}
Given an orientation rule $\mathcal D$ and undirected
cut-query access to $H$, the imbalance vector of the
induced orientation $D$ can be computed using $O(n)$
cut queries.
\end{observation}
\begin{proof}
For each vertex $v$, let
\[
 A_v=\{u:(v,u)\in\mathcal D\},
 \qquad
 B_v=\{u:(u,v)\in\mathcal D\}.
\]
By the definition of imbalance,
\[
 \bb_D(v)=c(\{v\},A_v)-c(\{v\},B_v).
\]
Each count can be computed using $O(1)$ queries by
\Cref{obs:edge-counting}, giving $O(n)$ queries in total.
\end{proof}

Finally, we make the following simple observation.
\begin{observation}[Skeleton cuts and imbalance]
\label{obs:skeleton-queries}
Let $G$ be an oriented graph and $H$ its undirected skeleton.
Then, for every $S\subseteq V$ and $v\in V$,
\[
 C_H(S)=F_G(S)+F_G(V\setminus S),
 \qquad
 \bb_G(v)=F_G(\{v\})-F_G(V\setminus\{v\}).
\]
Consequently, an undirected cut query to $H$ can be simulated
using two outgoing-cut queries to $G$, and the entire imbalance
vector $\bb_G$ can be computed using $2n$ outgoing-cut queries.
\end{observation}


\section{An Algorithmic Blueprint and the $\wtO(n^{9/5})$ Query Protocol}\label{sec:blueprint}

Our algorithm has three parts.
\begin{enumerate}
    \item \textbf{Reduction to constructing an orientation rule.}\label{bp:reduction}
    Let $H$ be the undirected skeleton of the input graph $G$,
    and let $\bb^*=\bb_G$. We show that every orientation of $H$
    with imbalance vector $\bb^*$ has the same reachability
    relation as $G$ (\Cref{lem:equivalence}). Thus we need not
    recover the original edge directions.

    Instead, we construct an \emph{orientation rule} $\mathcal D$
    that assigns a direction to every unordered pair of distinct
    vertices. Restricting this rule to the edges of $H$ gives
    its \emph{induced orientation} $D$. Our task is to construct
    a rule with $\bb_D=\bb^*$, without learning all the edges
    of $H$, and then solve reachability in $D$.

    \item \textbf{Approximating the target imbalance.} \label{bp:approx-imbalance}
    We first construct an orientation rule whose induced
    orientation $D$ has imbalance close to $\bb^*$ in $\ell_1$
    distance. Both the warm-up and the faster implementation use averaging and rounding
    for this step.


    \item \textbf{Repairing the remaining error by a flow.} \label{bp:flow-repair}
    Once $D$ is obtained, we give every edge of $D$ unit capacity and find an integral
    flow whose net outgoing flow at each vertex $v$ equals $(\bb_D(v)-\bb^*(v))/{2}$.
    We show that such a flow always exists and can be found with cut query access to $H$ (\Cref{lem:repair}). Reversing the edges
    carrying one unit of flow gives an orientation with
    imbalance exactly $\bb^*$. These reversals define a new
    orientation rule, whose induced orientation therefore
    preserves the reachability relation of $G$.
\end{enumerate}

In this section, we also explain how to implement the blueprint
using cut queries for the $\wtO(n^{9/5})$ warm-up algorithm.
The $\wtO(n^{5/3})$ algorithm uses the same reduction and flow
repair, but improves the averaging and rounding procedures used in the second step.
We defer its implementation to \S\ref{sec:improved}.

\paragraph{Comparison to \cite{CutReachability2026}.} Our approach shares a similar template with the earlier work \cite{CutReachability2026}. First, \cite{CutReachability2026} reduces the problem of $s$-$t$ reachability to computing a \emph{fractional} orientation rule $\bz$ inducing an imbalance vector $\bb_{\bz}$ with $\| \bb_{\bz} - \bb^*\|_1 \leq 1/\poly(n)$. Since $\bz$ need not be integral, it induces an edge weighting $H_{\bz}$ instead of an orientation. But, $\bb_{\bz}$ being such a good approximation of $\bb^*$ implies that cuts in $H_{\bz}$ have weight at least $1/2$ if the corresponding cut in $G$ has value at least $1$ and have weight at most $1/\poly(n)$ otherwise. In particular, this suffices for simulating BFS with cut queries and hence for $s$-$t$ reachability, \cite[Lemmas 3.2–3.4]{CutReachability2026}. This is a direct strengthening of the fact that we use, that an integral orientation inducing a matching imbalance vector suffices for $s$-$t$ reachability, \Cref{lem:equivalence}.

Given this reduction, the algorithm then computes such a fractional orientation rule $\bz$ in $\wtO(n^{1.5})$ queries, using an intricate cut-query implementation of an interior-point method based LP solver \cite{lee2019solving}. This step is direct in contrast with our two-phased approach.
\subsection{Reduction to constructing an orientation rule}
\label{sec:reduction}



We consider the following auxiliary problem.

\begin{definition}[Orientation rule with prescribed imbalances]
\label{defn:implicit-orient}
Given a simple undirected graph $H=(V,E)$ on $n$ vertices
and a feasible imbalance vector $\bb$, construct an orientation
rule $\mathcal D$ on $V$ whose induced orientation $D$ of $H$
satisfies $\bb_D=\bb$.
\end{definition}

We now mention the reachability-preservation property
underlying the reduction. This is already established in \cite{GhoshKMM26} (Corollary 3.2).

\begin{lemma}[Preserving imbalances preserves reachability]
\label{lem:equivalence}
Let $G$ be a directed graph with undirected skeleton $H$,
and let $\bb^*=\bb_G$. Every orientation $D$ of $H$ with
$\bb_D=\bb^*$ has the same outgoing-cut values and the same
reachability relation as $G$.
\end{lemma}
\begin{proof}
Fix $S\subseteq V$. Recall that $F_D(S)$ counts the edges
of $D$ directed from $S$ to $V\setminus S$. When the imbalances in $D$ are summed
over $S$, contributions from edges internal to $S$ cancel.
Each outgoing edge contributes $+1$ and each incoming edge
contributes $-1$. Since $C_H(S)$ counts both outgoing and
incoming edges, we have
\begin{equation}\label{eq:cut}
 2F_D(S)=C_H(S)+\sum_{v\in S}\bb_D(v).
\end{equation}
The same identity holds for $G$ with imbalance $\bb^*$.
Consequently, $\bb_D=\bb^*$ implies
$F_D(S)=F_G(S)$ for every $S\subseteq V$.

A vertex $t$ is unreachable from $s$ exactly when there is
a set containing $s$ but not $t$ with no outgoing edge.
In one direction, such a set prevents a path from reaching
$t$; in the other, take the set of all vertices reachable
from $s$. Equality of outgoing-cut values therefore
preserves reachability between every pair of vertices.
\end{proof}

By \Cref{lem:equivalence}, once we have constructed an orientation
rule $\mathcal D$ matching $\bb^*$ on $H$, it remains to solve
reachability in its induced orientation. We formalise this search
problem as follows.

\begin{definition}[Reachability under an orientation rule]
\label{def:reach-implicit}
The input consists of a simple undirected graph $H$ on
$n$ labelled vertices, an orientation rule $\mathcal D$ on
$V(H)$, and two vertices $s,t\in V(H)$. The task is to decide
whether $t$ is reachable from $s$ in the induced orientation
$D$ of $H$.
\end{definition}

In what follows, we show this problem can be solved using cut query access to $H$.




\subsubsection{Implementation Using Cut Queries}
\label{sec:reach-cut-implementation}

We first implement the graph search using undirected cut
queries to $H$. We then combine this with the construction
of an orientation rule to obtain the reduction. 

\begin{lemma}[Graph search under an orientation rule]
\label{lem:implicit-dfs}
Given an orientation rule $\mathcal D$ and undirected
cut-query access to $H$, the problem in
\Cref{def:reach-implicit} can be solved using $O(n\log n)$
queries.
\end{lemma}
\begin{proof}
We implement the out-neighbour searches used by the
graph-search procedure. Given a vertex $v$ and
a set $U$ containing $v$, form
\[
 X=\{w\in V\setminus U:(v,w)\in\mathcal D\}.
\]
This set can be computed from $\mathcal D$ and $U$ without
queries to $H$. The out-neighbours of $v$ outside $U$ in
$D$ are exactly the neighbours of $v$ in $X$ in $H$.
By \Cref{lem:cut-neighbour}, we can find one such neighbour,
or certify that none exists, using $O(\log n)$ queries. The graph search makes $O(n)$ such calls, giving
$O(n\log n)$ queries in total.
\end{proof}

We next reduce directed reachability to constructing an
orientation rule with prescribed imbalances
(\Cref{defn:implicit-orient}), followed by the search above. 

\begin{lemma}[Directed reachability via orientation rules]
\label{lem:directed-to-undirected}
Suppose the problem in \Cref{defn:implicit-orient} can be
solved using $Q(n)$ undirected cut queries on simple
undirected graphs with $n$ vertices. Then directed $s$-$t$
reachability on $n$ vertices can be solved using
$O(n\log n+Q(n))$ outgoing-cut queries.
\end{lemma}
\begin{proof}
Let $G$ be the input directed graph, let $H$ be its
undirected skeleton, and set $\bb^*=\bb_G$. The vector
$\bb^*$ is feasible for $H$, since $G$ itself is an
orientation of $H$ with this imbalance vector.

By \Cref{obs:skeleton-queries}, we can compute $\bb^*$ using
$2n$ outgoing-cut queries to $G$, and simulate each
undirected cut query to $H$ using two outgoing-cut queries.
Run the assumed algorithm on $H$ with target $\bb^*$ to
construct an orientation rule $\mathcal D$ matching $\bb^*$
on $H$. Computing $\bb^*$ and constructing $\mathcal D$
therefore use at most $2n+2Q(n)$ outgoing-cut queries.

Let $D$ be the induced orientation of $H$. By
\Cref{lem:equivalence}, $D$ has the same reachability
relation as $G$. We can therefore decide whether $t$ is
reachable from $s$ using \Cref{lem:implicit-dfs}. This takes
$O(n\log n)$ undirected cut queries to $H$, each simulated
by two outgoing-cut queries to $G$. The total number of
outgoing-cut queries is $O(n\log n+Q(n))$, as required.
\end{proof}

The rest of this section describes how to construct the
required orientation rule, first approximately (Step \ref{bp:approx-imbalance} of the blueprint) in \Cref{sec:meta}, and then
exactly (Step \ref{bp:flow-repair}) in \Cref{sec:repair}.

%% file: general-digraphs.tex
\section{Extension to General Unweighted Digraphs}
\label{app:general-digraphs}

We sketch how to remove the assumption that the input digraph is
oriented. Throughout this section, $G$ is an unweighted digraph, possibly with two arcs for a single unordered pair of vertices. Its undirected skeleton
$H$ is now a multigraph, with multiplicities $c_{uv}\in\{0,1,2\}$.
The two algorithms are essentially identical, retaining their respective query bounds
$\wtO(n^{9/5})$ and $\wtO(n^{5/3})$, in expectation and with high
probability. The main changes are to the representation of orientations
and the implementation of the cut-query primitives.

\subsection{Multi-orientation rules and the reduction from \Cref{sec:reduction}}

For every unordered pair $\{u,v\}$, fix two potential edge copies,
labelled $1$ and $2$ in order of priority. If the multiplicity of
the pair is $c_{uv}$, exactly the first $c_{uv}$ copies exist.
Thus copy $2$ cannot exist without copy $1$.

\begin{definition}[Multi-orientation rule]
\label{def:multi-orientation-rule}
A \emph{multi-orientation rule} $\mathcal D$ assigns a direction
to each of the two potential copies of every unordered pair of
distinct vertices. Its induced orientation $D$ of $H$ is obtained
by retaining the first $c_{uv}$ copies of each pair. We say that
$\mathcal D$ matches an imbalance vector $\bb$ if $\bb_D=\bb$,
where degrees count edge copies.
\end{definition}

Writing $\mathcal D(u,v,i)$ for the direction assigned to copy $i$
of $\{u,v\}$, we maintain the invariant
\begin{equation}\label{eq:multi-orientation-invariant}
 c_{uv}=1\quad\Longrightarrow\quad
 \mathcal D(u,v,1)=\mathcal D(u,v,2).
\end{equation}
Thus the nonexistent
second copy follows the first, without changing the induced
orientation. When both copies exist, their directions may differ.
The rule is specified simply by the two directions on every pair.
The algorithms below initially give both copies the same direction.
Whenever they change copies separately, they determine the pair's
multiplicity using $O(1)$ queries and preserve this invariant.

The invariant is needed for the query implementation. Without it,
a rule directing only copy $2$ out of a vertex $v$ could specify
an outgoing candidate even when the pair's sole existing copy
points into $v$. Under the invariant, a pair with an outgoing
candidate has an existing outgoing copy whenever $c_{uv}>0$.

A \emph{fractional multi-orientation rule} assigns one value
$\bz(u,v)\in[-1,1]$ to each pair $u<v$. Each existing copy has
probability $p_{uv}=(1+\bz(u,v))/2$ of pointing from $u$ to $v$.
Its imbalance is still given by
\eqref{eq:fractional-imbalance}, now with $c_{uv}\in\{0,1,2\}$.
In particular, a common fractional value suffices even if the
original arcs point in opposite directions. If $a_{uv}$ and
$a_{vu}$ count the original arcs in the two directions, then
\[
 \bz^*(u,v)=\frac{a_{uv}-a_{vu}}{c_{uv}}
 \qquad(c_{uv}>0)
\]
has imbalance $\bb_G$. Values on absent pairs are arbitrary.
The averaging procedures use this common fractional rule.
During rounding, the two copies may acquire different probabilities,
but we keep their probabilities equal whenever $c_{uv}=1$.

The analogue of \Cref{defn:implicit-orient} asks for a
multi-orientation rule satisfying invariant~\eqref{eq:multi-orientation-invariant}
and matching a feasible
imbalance vector on $H$. Once such a rule is constructed, we solve
the following analogue of \Cref{def:reach-implicit}.

\begin{definition}[Reachability under a multi-orientation rule]
\label{def:reach-multi-implicit}
The input consists of an undirected multigraph $H$ with
multiplicities at most two, a multi-orientation rule $\mathcal D$
satisfying invariant~\eqref{eq:multi-orientation-invariant}, and vertices
$s,t\in V(H)$. Given
undirected cut-query access to $H$, the task is to decide whether
$t$ is reachable from $s$ in the induced orientation $D$.
\end{definition}

We now show how to extend the reduction from \Cref{sec:reduction} to this more general setting. In fact, the proof of \Cref{lem:implicit-dfs} applies directly. To find an
out-neighbour of $v$ in a set $X$, restrict $X$ to vertices $w$
for which at least one potential copy of $\{v,w\}$ points out
of $v$. By the invariant, an existing pair in this candidate set
has an outgoing copy. We can therefore apply the same strategy as in
\Cref{lem:cut-neighbour}, the neighbor search subroutine. This search only distinguishes a zero
count from a positive count, so multiplicities do not affect it.
The resulting graph search still uses $O(n\log n)$ queries.

Finally, the cut identities used in \Cref{sec:reduction} actually already count
edge copies and therefore remain valid:
\[
 C_H(S)=F_G(S)+F_G(V\setminus S),
 \qquad
 2F_D(S)=C_H(S)+\sum_{v\in S}\bb_D(v).
\]
We can compute $\bb^*=\bb_G$ using $2n$ outgoing-cut queries,
and simulate an undirected cut query to $H$ using two outgoing-cut
queries. Matching $\bb^*$ preserves every outgoing-cut value and
hence reachability, exactly as in \Cref{lem:equivalence}.
Consequently, an algorithm constructing the required
multi-orientation rule in $Q(n)$ queries gives a directed
reachability algorithm using $O(Q(n)+n\log n)$ queries.

\subsection{Cut-query primitives on the multigraph}

As noted above, the edge-counting identity \eqref{eq:rectangle} returns the number
of copies between two sets. Thus \Cref{lem:cut-neighbour} still
finds a neighbour using $O(\log n)$ queries, and a pair's
multiplicity can be tested using $O(1)$ queries.

\paragraph{Integral imbalances.}
Let $A_v$ contain the vertices $w$ for which both potential
copies of $\{v,w\}$ point out of $v$, and let $B_v$ contain
those for which both point into $v$. By
invariant~\eqref{eq:multi-orientation-invariant}, every existing
pair with opposite directions has two copies, whose imbalance
contributions cancel. Hence
\[
 \bb_D(v)=c(\{v\},A_v)-c(\{v\},B_v).
\]
These two counts cost $O(1)$ queries by \Cref{obs:edge-counting},
so the whole imbalance vector costs $O(n)$ queries.

\paragraph{Fractional imbalances.}
More generally, suppose the two potential copies of a pair have
known coefficients $a_1,a_2\in[-1,1]$, with $a_1=a_2$ whenever
$c_{uv}=1$. Then their total contribution is
\[
 \sum_{i=1}^{c_{uv}}a_i=\frac{c_{uv}}2(a_1+a_2).
\]
We can therefore apply \Cref{lem:weighted-count} using the known
pair coefficient $(a_1+a_2)/2$. This computes approximate
fractional imbalances even when the copies have different
probabilities. There are at most $2(n-1)$ incident copies at a
vertex, so $O(\log n)$ binary digits still suffice for
inverse-polynomial additive error using $O(\log n)$ queries.

\subsection{The first two steps of the warm-up algorithm}

The averaging step of the warm-up directs both potential copies
from the smaller vertex value to the larger one. Its only query
access is through imbalance evaluations, which can still be computed, as discussed above. Since degrees are at most $2(n-1)$, the loss bounds change
only by a constant factor, as does the choice of step size. Hence, the resultant
average imbalance error remains $O(n^2/\sqrt T)$.

For independent rounding, choose one iteration independently for
each pair and give both copies its direction. Random choices are
independent between pairs, while the contribution of one pair to
a vertex imbalance has magnitude at most two. In particular,
\[
 \Var\bigl(\bb_D(v)\bigr)
 \leq\sum_{w\ne v}c_{vw}^2\leq4(n-1).
\]
This is the only change to the analysis, so the rounding error bound and its verification therefore have the
same query complexity as before.

\subsection{Repairing the multi-orientation rule using flow}
For repair, give each existing directed copy unit capacity.
The feasibility argument and the identity
$K=\|\bb_D-\bb^*\|_1/4$ from \Cref{lem:abstract-repair} are
unchanged. If there are two directed copies pointing the same direction, we combine the two arcs into one with capacity $2$.
Whenever a pair carries flow, determine its multiplicity and
update its two directions in the rule. An augmentation reverses copies
along the current residual orientation. Choose the required number
of existing copies in increasing priority order among those with
the required direction. In particular, a reversal on a pair of
multiplicity one must affect copy $1$, after which its nonexistent
second copy is assigned the same direction.

Internal residual-neighbour queries use the current multi-orientation
rule in exactly the same way as the graph search above. The total internal
residual capacity on an unordered pair is at most two. Thus the
layer argument in \Cref{lem:remaining-flow-distance} gives
$R\leq8n^2/h^2$ instead of $R\leq4n^2/h^2$. However, this only leads to the remainder of
the analysis in Appendix~\ref{app:flow} changing by constants, yielding the same 
$\wtO(n^{5/3}+n\sqrt{\|\bb_D-\bb^*\|_1})$ query bound as before.
\subsection{The faster algorithm}

The averaging procedure in \Cref{sec:fast-average} maintains
one fractional value for each pair. In the analysis, use one
coordinate for each existing copy; the coordinates belonging to
the same pair evolve identically. There are still $O(n^2)$
coordinates, and the degree bound is now $2(n-1)$. The regret
and Lipschitz estimates consequently change only by constants, and we can adjust the constant in the step size accordingly.
Approximate weighted imbalance evaluations are the only query
access needed by this procedure, and this can be implemented via the multigraph version of
\Cref{lem:weighted-count} described above. Thus faster averaging
still costs $O(nT\log n)$ queries and gives error $O(n^2/T)$.

It remains to extend the balanced-orientation construction used
in dependent rounding. The Eulerian-tour argument and the
completion lemma in Appendix~\ref{app:balanced-completion} apply
to edge copies without change. When constructing successive
spanning forests, learn each encountered pair's multiplicity and
remove its chosen copies explicitly. For the additional random
reserve, sample potential pairs independently and reserve all
remaining copies of each sampled pair. Orient the nonreserved
copies using one independent random direction per pair.

These changes preserve the query cost of constructing the
reserve. They also preserve the concentration argument up to
constants: if a cut has $a$ copies, its independent pair
contributions have magnitudes $m_{uv}\leq2$, with
\[
 \sum_{\{u,v\}\text{ crossing the cut}}m_{uv}^2
 \leq2\sum_{\{u,v\}\text{ crossing the cut}}m_{uv}=2a.
\]
The cut-counting argument therefore gives the same high-probability
completion guarantee after adjusting constants. The reserved
pairs have known multiplicities, so completion can orient their
copies separately while giving any nonexistent second copy the
direction of the first. This proves the multigraph version
of \Cref{lem:balanced-2} with the same query bound.

Finally, apply the bit-by-bit rounding procedure of
\Cref{lem:dependent-round}. The two potential copies may have
different probabilities, but their probabilities agree whenever
$c_{uv}=1$. Their membership in each bit subgraph therefore also
agrees on such pairs. Each balanced orientation gives the
nonexistent second copy the direction of the first, so the two
probabilities remain equal after the bit update. Bit subgraphs
retain their original copy labels and use cut queries to $H$. When we assign different directions to the edges between some pair $\{u,v\}$, we retain the knowledge that $c_{uv} = 2$. We use this retained information to correct the cut queries
used in neighbor searches and imbalance evaluations.
The primitives above
therefore implement each call to the balanced-orientation
subroutine and preserve the invariant for the next bit.
The initial truncation error remains $O(n2^{-B})$ per vertex,
and each bit changes the imbalance by $2^{1-j}\bb_{D_j}$ as
before. Hence dependent rounding still has $O(1)$ error per
vertex and uses $\wtO(n^{5/3})$ total queries. Combining this with
faster averaging, the flow repair step, and $T=n^{2/3}$
gives the $\wtO(n^{5/3})$ bound for general unweighted digraphs.

%% file: biblio.bib
@inproceedings{JiangNS26,
  author       = {Yonggang Jiang and Danupon Nanongkai and Pachara Sawettamalya},
  title        = {Minimum {$s$--$t$} Cuts with Fewer Cut Queries},
  booktitle    = {Proceedings of the 2026 ACM-SIAM Symposium on Discrete Algorithms ({SODA})},
  pages        = {258--296},
  publisher    = {{SIAM}},
  year         = {2026},
  doi          = {10.1137/1.9781611978971.12},
  note         = {\href{https://doi.org/10.1137/1.9781611978971.12}{doi:\nolinkurl{10.1137/1.9781611978971.12}}},
}

@inproceedings{GhoshKMM26,
  author       = {Prantar Ghosh and Sahil Kuchlous and Shravan Mehra and Sagnik Mukhopadhyay},
  title        = {The Power of the Score Sequence of a Tournament},
  booktitle    = {European Symposium on Algorithms ({ESA} 2026)},
  series       = {Leibniz International Proceedings in Informatics ({LIPIcs})},
  volume       = {388},
  pages        = {156:1--156:19},
  publisher    = {Schloss Dagstuhl -- Leibniz-Zentrum f{\"u}r Informatik},
  year         = {2026},
  doi          = {10.4230/LIPIcs.ESA.2026.156},
  note         = {\href{https://doi.org/10.4230/LIPIcs.ESA.2026.156}{doi:\nolinkurl{10.4230/LIPIcs.ESA.2026.156}}},
}

@misc{CutReachability2026,
  title        = {Directed Reachability from Cut Queries in {$\tilde O(n^{3/2})$} Queries},
  year         = {2026},
  howpublished = {Unpublished manuscript},
  url          = {https://yangpliu.github.io/repo/directed-reachability-cut-queries/paper.pdf},
  note         = {\url{https://yangpliu.github.io/repo/directed-reachability-cut-queries/paper.pdf}},
  key          = {Directed Reachability},
}

@article{EneN2026,
  author       = {Alina Ene and Huy L. Nguyen},
  title        = {A Faster Strongly Polynomial Time Algorithm for Submodular Function Minimization},
  journal      = {arXiv preprint arXiv:2610.25403},
  year         = {2026},
  eprint       = {2610.25403},
  archiveprefix = {arXiv},
  url          = {https://arxiv.org/abs/2610.25403},
  note         = {\href{https://arxiv.org/abs/2610.25403}{arXiv:2610.25403}},
}

@article{KhannaPS26,
  author       = {Sanjeev Khanna and Aaron Putterman and Junkai Song},
  title        = {Reachability in Directed Acyclic Graphs with Near-Linear Cut Queries},
  journal      = {arXiv preprint arXiv:2607.21390},
  year         = {2026},
  eprint       = {2607.21390},
  archiveprefix = {arXiv},
  url          = {https://arxiv.org/abs/2607.21390},
  note         = {\href{https://arxiv.org/abs/2607.21390}{arXiv:2607.21390}},
}

@article{BalsTN26,
  author       = {Ben Bals and Matei Tinca and Yasamin Nazari},
  title        = {Cut Query Reachability for {DAGs} with Subquadratic Queries},
  journal      = {arXiv preprint arXiv:2607.26630},
  year         = {2026},
  eprint       = {2607.26630},
  archiveprefix = {arXiv},
  url          = {https://arxiv.org/abs/2607.26630},
  note         = {\href{https://arxiv.org/abs/2607.26630}{arXiv:2607.26630}},
}

@inproceedings{Kenneth-Mordoch26b,
  author       = {Yotam Kenneth{-}Mordoch and Robert Krauthgamer},
  title        = {Faster All-Pairs Minimum Cut: Bypassing Exact Max-Flow},
  booktitle    = {Proceedings of the 2026 ACM Symposium on Theory of Computing ({STOC})},
  pages        = {1254--1265},
  publisher    = {{ACM}},
  year         = {2026},
  doi          = {10.1145/3798129.3800836},
  note         = {\href{https://doi.org/10.1145/3798129.3800836}{doi:\nolinkurl{10.1145/3798129.3800836}}. Full version: \href{https://arxiv.org/abs/2511.10036}{arXiv:2511.10036}},
}

@inproceedings{Kenneth-Mordoch26,
  author       = {Yotam Kenneth{-}Mordoch and Robert Krauthgamer},
  title        = {All-Pairs Minimum Cut using {$\widetilde O(n^{7/4})$} Cut Queries},
  booktitle    = {Proceedings of the 2026 ACM-SIAM Symposium on Discrete Algorithms ({SODA})},
  pages        = {4077--4095},
  publisher    = {{SIAM}},
  year         = {2026},
}

@inproceedings{ChakrabartyGJS22,
  author       = {Deeparnab Chakrabarty and Andrei Graur and Haotian Jiang and Aaron Sidford},
  title        = {Improved Lower Bounds for Submodular Function Minimization},
  booktitle    = {Proceedings of the 2022 IEEE Symposium on Foundations of Computer Science ({FOCS})},
  pages        = {245--254},
  publisher    = {{IEEE} Computer Society},
  year         = {2022},
  doi          = {10.1109/FOCS54457.2022.00030},
  note         = {\href{https://doi.org/10.1109/FOCS54457.2022.00030}{doi:\nolinkurl{10.1109/FOCS54457.2022.00030}}},
}

@inproceedings{ApersEGLMN22,
  author       = {Simon Apers and Yuval Efron and Pawel Gawrychowski and Troy Lee and Sagnik Mukhopadhyay and Danupon Nanongkai},
  title        = {Cut Query Algorithms with Star Contraction},
  booktitle    = {Proceedings of the 2022 IEEE Symposium on Foundations of Computer Science ({FOCS})},
  pages        = {507--518},
  publisher    = {{IEEE} Computer Society},
  year         = {2022},
  doi          = {10.1109/FOCS54457.2022.00055},
  note         = {\href{https://doi.org/10.1109/FOCS54457.2022.00055}{doi:\nolinkurl{10.1109/FOCS54457.2022.00055}}},
}

@inproceedings{blikstadBEMN22,
  author       = {Joakim Blikstad and Jan van den Brand and Yuval Efron and Sagnik Mukhopadhyay and Danupon Nanongkai},
  title        = {Nearly Optimal Communication and Query Complexity of Bipartite Matching},
  booktitle    = {Proceedings of the 2022 IEEE Symposium on Foundations of Computer Science ({FOCS})},
  pages        = {1174--1185},
  publisher    = {{IEEE} Computer Society},
  year         = {2022},
  doi          = {10.1109/FOCS54457.2022.00113},
  note         = {\href{https://doi.org/10.1109/FOCS54457.2022.00113}{doi:\nolinkurl{10.1109/FOCS54457.2022.00113}}},
}

@inproceedings{RubinsteinSW18,
  author       = {Aviad Rubinstein and Tselil Schramm and S. Matthew Weinberg},
  title        = {Computing Exact Minimum Cuts Without Knowing the Graph},
  booktitle    = {Innovations in Theoretical Computer Science Conference ({ITCS} 2018)},
  series       = {Leibniz International Proceedings in Informatics ({LIPIcs})},
  volume       = {94},
  pages        = {39:1--39:16},
  publisher    = {Schloss Dagstuhl -- Leibniz-Zentrum f{\"u}r Informatik},
  year         = {2018},
  doi          = {10.4230/LIPIcs.ITCS.2018.39},
  url          = {https://doi.org/10.4230/LIPIcs.ITCS.2018.39},
  note         = {\href{https://doi.org/10.4230/LIPIcs.ITCS.2018.39}{doi:\nolinkurl{10.4230/LIPIcs.ITCS.2018.39}}},
}

@inproceedings{LeeSW15,
  author       = {Yin Tat Lee and Aaron Sidford and Sam Chiu{-}wai Wong},
  title        = {A Faster Cutting Plane Method and its Implications for Combinatorial and Convex Optimization},
  booktitle    = {Proceedings of the 2015 IEEE Symposium on Foundations of Computer Science ({FOCS})},
  pages        = {1049--1065},
  publisher    = {{IEEE} Computer Society},
  year         = {2015},
  doi          = {10.1109/FOCS.2015.68},
  url          = {https://doi.org/10.1109/FOCS.2015.68},
  note         = {\href{https://doi.org/10.1109/FOCS.2015.68}{doi:\nolinkurl{10.1109/FOCS.2015.68}}},
}

@inproceedings{MukhopadhyayN20,
  author       = {Sagnik Mukhopadhyay and Danupon Nanongkai},
  title        = {Weighted min-cut: sequential, cut-query, and streaming algorithms},
  booktitle    = {Proceedings of the 2020 ACM Symposium on Theory of Computing ({STOC})},
  pages        = {496--509},
  publisher    = {{ACM}},
  year         = {2020},
  doi          = {10.1145/3357713.3384334},
  url          = {https://doi.org/10.1145/3357713.3384334},
  note         = {\href{https://doi.org/10.1145/3357713.3384334}{doi:\nolinkurl{10.1145/3357713.3384334}}},
  _editor      = {Konstantin Makarychev and Yury Makarychev and Madhur Tulsiani and Gautam Kamath and Julia Chuzhoy},
}

@article{Jiang23,
  author       = {Haotian Jiang},
  title        = {Minimizing Convex Functions with Rational Minimizers},
  journal      = {Journal of the ACM},
  volume       = {70},
  number       = {1},
  pages        = {5:1--5:27},
  year         = {2022},
  doi          = {10.1145/3566050},
  note         = {\href{https://doi.org/10.1145/3566050}{doi:\nolinkurl{10.1145/3566050}}},
}

@inproceedings{AxelrodLS20,
  author       = {Brian Axelrod and Yang P. Liu and Aaron Sidford},
  title        = {Near-optimal Approximate Discrete and Continuous Submodular Function Minimization},
  booktitle    = {Proceedings of the 2020 ACM-SIAM Symposium on Discrete Algorithms ({SODA})},
  pages        = {837--853},
  publisher    = {{SIAM}},
  year         = {2020},
  doi          = {10.1137/1.9781611975994.51},
  note         = {\href{https://doi.org/10.1137/1.9781611975994.51}{doi:\nolinkurl{10.1137/1.9781611975994.51}}},
}

@article{Vygen03,
  author       = {Jens Vygen},
  title        = {A note on {Schrijver}'s submodular function minimization algorithm},
  journal      = {Journal of Combinatorial Theory, Series B},
  volume       = {88},
  number       = {2},
  pages        = {399--402},
  year         = {2003},
  doi          = {10.1016/S0095-8956(02)00047-3},
  note         = {\href{https://doi.org/10.1016/S0095-8956(02)00047-3}{doi:\nolinkurl{10.1016/S0095-8956(02)00047-3}}},
}

@article{Orlin09,
  author       = {James B. Orlin},
  title        = {A faster strongly polynomial time algorithm for submodular function minimization},
  journal      = {Mathematical Programming},
  volume       = {118},
  number       = {2},
  pages        = {237--251},
  year         = {2009},
  doi          = {10.1007/s10107-007-0189-2},
  url          = {https://doi.org/10.1007/s10107-007-0189-2},
  note         = {\href{https://doi.org/10.1007/s10107-007-0189-2}{doi:\nolinkurl{10.1007/s10107-007-0189-2}}},
}

@inproceedings{IwataO09,
  author       = {Satoru Iwata and James B. Orlin},
  title        = {A simple combinatorial algorithm for submodular function minimization},
  booktitle    = {Proceedings of the 2009 ACM-SIAM Symposium on Discrete Algorithms ({SODA})},
  pages        = {1230--1237},
  publisher    = {{SIAM}},
  year         = {2009},
  doi          = {10.1137/1.9781611973068.133},
  note         = {\href{https://doi.org/10.1137/1.9781611973068.133}{doi:\nolinkurl{10.1137/1.9781611973068.133}}},
}

@article{Iwata03,
  author       = {Satoru Iwata},
  title        = {A Faster Scaling Algorithm for Minimizing Submodular Functions},
  journal      = {SIAM Journal on Computing},
  volume       = {32},
  number       = {4},
  pages        = {833--840},
  year         = {2003},
  doi          = {10.1137/S0097539701397813},
  note         = {\href{https://doi.org/10.1137/S0097539701397813}{doi:\nolinkurl{10.1137/S0097539701397813}}},
}

@article{Schrijver00,
  author       = {Alexander Schrijver},
  title        = {A Combinatorial Algorithm Minimizing Submodular Functions in Strongly Polynomial Time},
  journal      = {Journal of Combinatorial Theory, Series B},
  volume       = {80},
  number       = {2},
  pages        = {346--355},
  year         = {2000},
  doi          = {10.1006/jctb.2000.1989},
  url          = {https://doi.org/10.1006/jctb.2000.1989},
  note         = {\href{https://doi.org/10.1006/jctb.2000.1989}{doi:\nolinkurl{10.1006/jctb.2000.1989}}},
}

@article{IwataFF01,
  author       = {Satoru Iwata and Lisa Fleischer and Satoru Fujishige},
  title        = {A combinatorial strongly polynomial algorithm for minimizing submodular functions},
  journal      = {Journal of the ACM},
  volume       = {48},
  number       = {4},
  pages        = {761--777},
  year         = {2001},
  doi          = {10.1145/502090.502096},
  url          = {https://doi.org/10.1145/502090.502096},
  note         = {\href{https://doi.org/10.1145/502090.502096}{doi:\nolinkurl{10.1145/502090.502096}}},
}

@article{FleischerI03,
  author       = {Lisa Fleischer and Satoru Iwata},
  title        = {A push-relabel framework for submodular function minimization and applications to parametric optimization},
  journal      = {Discrete Applied Mathematics},
  volume       = {131},
  number       = {2},
  pages        = {311--322},
  year         = {2003},
  doi          = {10.1016/S0166-218X(02)00458-4},
  note         = {\href{https://doi.org/10.1016/S0166-218X(02)00458-4}{doi:\nolinkurl{10.1016/S0166-218X(02)00458-4}}},
}

@article{Cunningham85,
  author       = {William H. Cunningham},
  title        = {On submodular function minimization},
  journal      = {Combinatorica},
  volume       = {5},
  number       = {3},
  pages        = {185--192},
  year         = {1985},
  doi          = {10.1007/BF02579361},
  note         = {\href{https://doi.org/10.1007/BF02579361}{doi:\nolinkurl{10.1007/BF02579361}}},
}

@book{GLS1988,
  author       = {Martin Gr{\"{o}}tschel and L{\'{a}}szl{\'{o}} Lov{\'{a}}sz and Alexander Schrijver},
  title        = {Geometric Algorithms and Combinatorial Optimization},
  series       = {Algorithms and Combinatorics},
  volume       = {2},
  publisher    = {Springer},
  year         = {1988},
  doi          = {10.1007/978-3-642-97881-4},
  note         = {\href{https://doi.org/10.1007/978-3-642-97881-4}{doi:\nolinkurl{10.1007/978-3-642-97881-4}}},
  isbn         = {978-3-642-97883-8},
}

@article{GrotschelLS81,
  author       = {Martin Gr{\"{o}}tschel and L{\'{a}}szl{\'{o}} Lov{\'{a}}sz and Alexander Schrijver},
  title        = {The ellipsoid method and its consequences in combinatorial optimization},
  journal      = {Combinatorica},
  volume       = {1},
  number       = {2},
  pages        = {169--197},
  year         = {1981},
  doi          = {10.1007/BF02579273},
  note         = {\href{https://doi.org/10.1007/BF02579273}{doi:\nolinkurl{10.1007/BF02579273}}},
}

@incollection{Lovasz82,
  author       = {L{\'{a}}szl{\'{o}} Lov{\'{a}}sz},
  title        = {Submodular functions and convexity},
  editor       = {Achim Bachem and Bernhard Korte and Martin Gr{\"o}tschel},
  booktitle    = {Mathematical Programming: The State of the Art},
  pages        = {235--257},
  publisher    = {Springer},
  year         = {1983},
  doi          = {10.1007/978-3-642-68874-4_10},
  note         = {\href{https://doi.org/10.1007/978-3-642-68874-4_10}{doi:\nolinkurl{10.1007/978-3-642-68874-4_10}}},
}

@inproceedings{ChakrabartyLS0W19,
  author       = {Deeparnab Chakrabarty and Yin Tat Lee and Aaron Sidford and Sahil Singla and Sam Chiu{-}wai Wong},
  title        = {Faster Matroid Intersection},
  booktitle    = {Proceedings of the 2019 IEEE Symposium on Foundations of Computer Science ({FOCS})},
  pages        = {1146--1168},
  publisher    = {{IEEE} Computer Society},
  year         = {2019},
  doi          = {10.1109/FOCS.2019.00072},
  note         = {\href{https://doi.org/10.1109/FOCS.2019.00072}{doi:\nolinkurl{10.1109/FOCS.2019.00072}}},
}

@inproceedings{Jiang21,
  author       = {Haotian Jiang},
  title        = {Minimizing Convex Functions with Integral Minimizers},
  booktitle    = {Proceedings of the 2021 ACM-SIAM Symposium on Discrete Algorithms ({SODA})},
  pages        = {976--985},
  publisher    = {{SIAM}},
  year         = {2021},
  doi          = {10.1137/1.9781611976465.61},
  note         = {\href{https://doi.org/10.1137/1.9781611976465.61}{doi:\nolinkurl{10.1137/1.9781611976465.61}}},
}

@inproceedings{Harvey08,
  author       = {Nicholas J. A. Harvey},
  title        = {Matroid intersection, pointer chasing, and {Young}'s seminormal representation of {$S_n$}},
  booktitle    = {Proceedings of the 2008 ACM-SIAM Symposium on Discrete Algorithms ({SODA})},
  pages        = {542--549},
  publisher    = {{SIAM}},
  year         = {2008},
  url          = {http://dl.acm.org/citation.cfm?id=1347082.1347142},
}

@article{lee2019solving,
  author       = {Lee, Yin Tat and Sidford, Aaron},
  title        = {Solving linear programs with {$\sqrt{\operatorname{rank}}$} linear system solves},
  journal      = {arXiv preprint arXiv:1910.08033},
  year         = {2019},
  eprint       = {1910.08033},
  archiveprefix = {arXiv},
  url          = {https://arxiv.org/abs/1910.08033},
  note         = {\href{https://arxiv.org/abs/1910.08033}{arXiv:1910.08033}},
}

@article{ET75,
  title={Network flow and testing graph connectivity},
  author={Even, Shimon and Tarjan, R Endre},
  journal={SIAM journal on computing},
  volume={4},
  number={4},
  pages={507--518},
  year={1975},
  publisher={SIAM}
}

@inproceedings{blumenstock2016fast,
  title={Fast algorithms for pseudoarboricity},
  author={Blumenstock, Markus},
  booktitle={2016 Proceedings of the Eighteenth Workshop on Algorithm Engineering and Experiments (ALENEX)},
  pages={113--126},
  year={2016},
  organization={SIAM}
}

@book{cormen1990introduction,
  title={Introduction to algorithms},
  author={Cormen, Thomas H and Leiserson, Charles Eric and Rivest, Ronald L and Stein, Clifford},
  volume={3},
  year={2009},
  publisher={MIT press, Cambridge}
}


%% file: references.bib
@inproceedings{ASW,
  author       = {Anand, Aditya and Saranurak, Thatchaphol and Wang, Yunfan},
  title        = {Deterministic edge connectivity and max flow using subquadratic cut queries},
  booktitle    = {Proceedings of the 2025 ACM-SIAM Symposium on Discrete Algorithms ({SODA})},
  pages        = {124--142},
  publisher    = {{SIAM}},
  year         = {2025},
  doi          = {10.1137/1.9781611978322.4},
  note         = {\href{https://doi.org/10.1137/1.9781611978322.4}{doi:\nolinkurl{10.1137/1.9781611978322.4}}. Full version: \href{https://arxiv.org/abs/2410.18704}{arXiv:2410.18704}},
}

@article{Nemirovski,
  author       = {Nemirovski, Arkadi},
  title        = {Prox-method with rate of convergence {$O(1/t)$} for variational inequalities with {Lipschitz} continuous monotone operators and smooth convex-concave saddle point problems},
  journal      = {SIAM Journal on Optimization},
  volume       = {15},
  number       = {1},
  pages        = {229--251},
  publisher    = {SIAM},
  year         = {2004},
  doi          = {10.1137/S1052623403425629},
  note         = {\href{https://doi.org/10.1137/S1052623403425629}{doi:\nolinkurl{10.1137/S1052623403425629}}},
}

@article{BK,
  author       = {Andr{\'a}s A. Bencz{\'u}r and David R. Karger},
  title        = {Randomized approximation schemes for cuts and flows in capacitated graphs},
  journal      = {SIAM Journal on Computing},
  volume       = {44},
  number       = {2},
  pages        = {290--319},
  publisher    = {SIAM},
  year         = {2015},
  doi          = {10.1137/070705970},
  note         = {\href{https://doi.org/10.1137/070705970}{doi:\nolinkurl{10.1137/070705970}}},
}

@incollection{D06,
  author       = {Dinitz, Yefim},
  title        = {{Dinitz}' algorithm: The original version and {Even}'s version},
  editor       = {Oded Goldreich and Arnold L. Rosenberg and Alan L. Selman},
  booktitle    = {Theoretical Computer Science: Essays in Memory of Shimon Even},
  series       = {Lecture Notes in Computer Science},
  volume       = {3895},
  pages        = {218--240},
  publisher    = {Springer},
  year         = {2006},
  doi          = {10.1007/11685654_10},
  note         = {\href{https://doi.org/10.1007/11685654_10}{doi:\nolinkurl{10.1007/11685654_10}}},
}

@article{KS96,
  author       = {Karger, David R. and Stein, Clifford},
  title        = {A new approach to the minimum cut problem},
  journal      = {Journal of the ACM},
  volume       = {43},
  number       = {4},
  pages        = {601--640},
  year         = {1996},
  doi          = {10.1145/234533.234534},
  note         = {\href{https://doi.org/10.1145/234533.234534}{doi:\nolinkurl{10.1145/234533.234534}}. Author version: \href{https://people.csail.mit.edu/karger/Papers/contract.pdf}{https://people.csail.mit.edu/karger/Papers/contract.pdf}},
}

@inproceedings{RS13,
  author       = {Rakhlin, Alexander and Sridharan, Karthik},
  title        = {Optimization, learning, and games with predictable sequences},
  booktitle    = {Advances in Neural Information Processing Systems 26},
  year         = {2013},
  note         = {\href{https://proceedings.neurips.cc/paper_files/paper/2013/file/f0dd4a99fba6075a9494772b58f95280-Paper.pdf}{Proceedings paper}. Full version: \href{https://www.cs.cornell.edu/~sridharan/predgamesopt.pdf}{https://www.cs.cornell.edu/~sridharan/predgamesopt.pdf}},
}

@article{SS12,
  author       = {Shalev-Shwartz, Shai},
  title        = {Online learning and online convex optimization},
  journal      = {Foundations and Trends in Machine Learning},
  volume       = {4},
  number       = {2},
  pages        = {107--194},
  year         = {2012},
  doi          = {10.1561/2200000018},
  note         = {\href{https://doi.org/10.1561/2200000018}{doi:\nolinkurl{10.1561/2200000018}}},
}

@misc{Trevisan19,
  author       = {Trevisan, Luca},
  title        = {Online optimization post 3: Follow the regularized leader},
  year         = {2019},
  howpublished = {\emph{in theory}, May 6},
  note         = {\href{https://lucatrevisan.wordpress.com/2019/05/06/online-optimization-post-3-follow-the-regularized-leader/}{https://lucatrevisan.wordpress.com/2019/05/06/online-optimization-post-3-follow-the-regularized-leader/}},
}

@inproceedings{bodwin23,
  author       = {Bodwin, Greg and Fleischmann, Henry},
  title        = {Spanning Adjacency Oracles in Sublinear Time},
  editor       = {Guruswami, Venkatesan},
  booktitle    = {Innovations in Theoretical Computer Science Conference ({ITCS} 2024)},
  series       = {Leibniz International Proceedings in Informatics ({LIPIcs})},
  volume       = {287},
  pages        = {19:1--19:21},
  publisher    = {Schloss Dagstuhl -- Leibniz-Zentrum f{\"u}r Informatik},
  address      = {Dagstuhl, Germany},
  year         = {2024},
  doi          = {10.4230/LIPIcs.ITCS.2024.19},
  url          = {https://drops.dagstuhl.de/entities/document/10.4230/LIPIcs.ITCS.2024.19},
  note         = {\href{https://doi.org/10.4230/LIPIcs.ITCS.2024.19}{doi:\nolinkurl{10.4230/LIPIcs.ITCS.2024.19}}},
  isbn         = {978-3-95977-309-6},
  issn         = {1868-8969},
  urn          = {urn:nbn:de:0030-drops-195475},
}

@article{nagamochi1992linear,
  author       = {Nagamochi, Hiroshi and Ibaraki, Toshihide},
  title        = {A linear-time algorithm for finding a sparse {$k$}-connected spanning subgraph of a {$k$}-connected graph},
  journal      = {Algorithmica},
  volume       = {7},
  number       = {1},
  pages        = {583--596},
  publisher    = {Springer},
  year         = {1992},
  doi          = {10.1007/BF01758778},
  note         = {\href{https://doi.org/10.1007/BF01758778}{doi:\nolinkurl{10.1007/BF01758778}}},
}
